\documentclass[british,aps]{revtex4}
\usepackage[T1]{fontenc}
\usepackage[utf8]{inputenc}
\usepackage[a4paper]{geometry}
\usepackage{float}
\usepackage{textcomp}
\usepackage{amstext}
\usepackage{graphicx}

\makeatletter

\@ifundefined{textcolor}{}
{%
 \definecolor{BLACK}{gray}{0}
 \definecolor{WHITE}{gray}{1}
 \definecolor{RED}{rgb}{1,0,0}
 \definecolor{GREEN}{rgb}{0,1,0}
 \definecolor{BLUE}{rgb}{0,0,1}
 \definecolor{CYAN}{cmyk}{1,0,0,0}
 \definecolor{MAGENTA}{cmyk}{0,1,0,0}
 \definecolor{YELLOW}{cmyk}{0,0,1,0}
 }

\makeatother

\usepackage{babel}

\begin{document}

\title{In Situ Characterisation of Permanent Magnetic Quadrupoles for Focussing
Proton Beams}

\author{J.J. Melone$^{13}$, K.W.D. Ledingham$^{123}$, T. McCanny$^{1}$,
T. Burris-Mog$^{2}$, U. Schramm$^{2}$,\\
 R. Grötzschel$^{2}$, S. Akhmadaliev$^{2}$, D. Hanf$^{2}$, K.M.
Spohr$^{4}$, \\
M. Bussmann$^{2}$ , T. Cowan$^{2}$ , S. M. Wiggins$^{1}$, M.
R. Mitchell$^{\text{1}}$ }

\affiliation{$^{1}$SUPA Department of Physics, University of Strathclyde, Glasgow
G4 0NG, United Kingdom\\
$^{2}$Institut für Strahlenphysik, Forschungzentrum Dresden-Rossendorf,
Bautzner Landstraße 128, 01328 Dresden, Germany\\
$^{3}$AWE plc., Aldermaston, Reading RG7 4PR, United Kingdom\\
$^{4}$SUPA Department of Physics, University of the West of Scotland,
Paisley PA1 2BE, United Kingdom}
\begin{abstract}
High intensity laser driven proton beams are at present receiving
much attention. The reasons for this are many but high on the list
is the potential to produce compact accelerators. However two of the
limitations of this technology is that unlike conventional nuclear
RF accelerators lasers produce diverging beams with an exponential
energy distribution. A number of different approaches have been attempted
to monochromise these beams but it has become obvious that magnetic
spectrometer technology developed over many years by nuclear physicists
to transport and focus proton beams could play an important role for
this purpose. This paper deals with the design and characterisation
of a magnetic quadrupole system which will attempt to focus and transport
laser-accelerated proton beams.
\end{abstract}
\maketitle

\section{Introduction}

Outstanding progress has been made in high-power laser technology
in the last 10 years with laser powers reaching petawatt (PW) values.
Petawatt lasers generate electric fields of 10$^{12}$ Vm$^{-1}$
with a large fraction of the total pulse energy being converted to
relativistic electrons with energies reaching in excess of 1 GeV.
In turn these electrons can result in the generation of beams of protons,
heavy ions, neutrons and high-energy photons. These laser-driven particle
beams have encouraged many to think of carrying out experiments normally
associated with conventional nuclear accelerators and reactors. To
this end a number of introductory articles have been written under
a trial name \textquoteleft{}Laser Nuclear Physics\textquoteright{}
\cite{ledingham_norreys99,Hideaki,ledingham_mckenna2003,ledingham-galster}.

However two of the draw backs for this technology are that unlike
conventional nuclear RF accelerators, lasers produce diverging ion
beams with exponential energy distributions. A number of different
approaches have been attempted to monochromise these beams using sophisticated
shaped targetry\cite{schwoerer}, specially controlled chemical treatment
of high Z metal foils\cite{hegelich} and the use of micro-lens technology
requiring a second laser beam\cite{Toncian}. These approaches have
had some success at the expense of increased complication.

In a recent paper entitled \textquotedblleft{}What will it take for
laser driven proton accelerators to be applied to tumour therapy\textquotedblright{}\cite{linz}
the authors correctly pointed out that conventional nuclear magnetic
transport systems could be used for transporting particles from laser
accelerators with the correct energy characteristics to wherever they
are needed. However they also point out that this might defeat the
basic premise of compactness of the new laser technology. 

On the contrary, however, over the last few years very compact permanent
magnet systems using neodymium iron boron $(Nd_{2}Fe_{14}B)$ alloy\cite{neomagnets}
sectors in a Halbach geometry\cite{Halbach} have been investigated
as suitable magnetic spectrometers particularly with laser based systems
in mind. These have been developed as miniature magnetic devices for
laser based accelerators particularly to control electron beams for
the production of undulator or free electron radiation\cite{wiggins_plasma}
or inverse-Compton scattering production of X-rays\cite{jklim,eichner,chen,schollmeier,becker}.
In particular Schollmeier et al.\cite{schollmeier} emphasise that
two permanent quadrupoles (PMQ) transport and focus laser accelerated
protons in a very reproducible and predictable manner decoupling the
acceleration process from the beam transport allowing for independent
optimisation of the processes.

The present paper will describe the design and characterisation of
a Halbach system particularly designed for laser produced protons
up to about 20 MeV. However for characterisation of the system protons
up to 9 MeV from a tandem were used.

\section{Experimental Set-up}

\begin{figure}[H]
\begin{centering}
(a)\includegraphics[width=0.3\paperwidth]{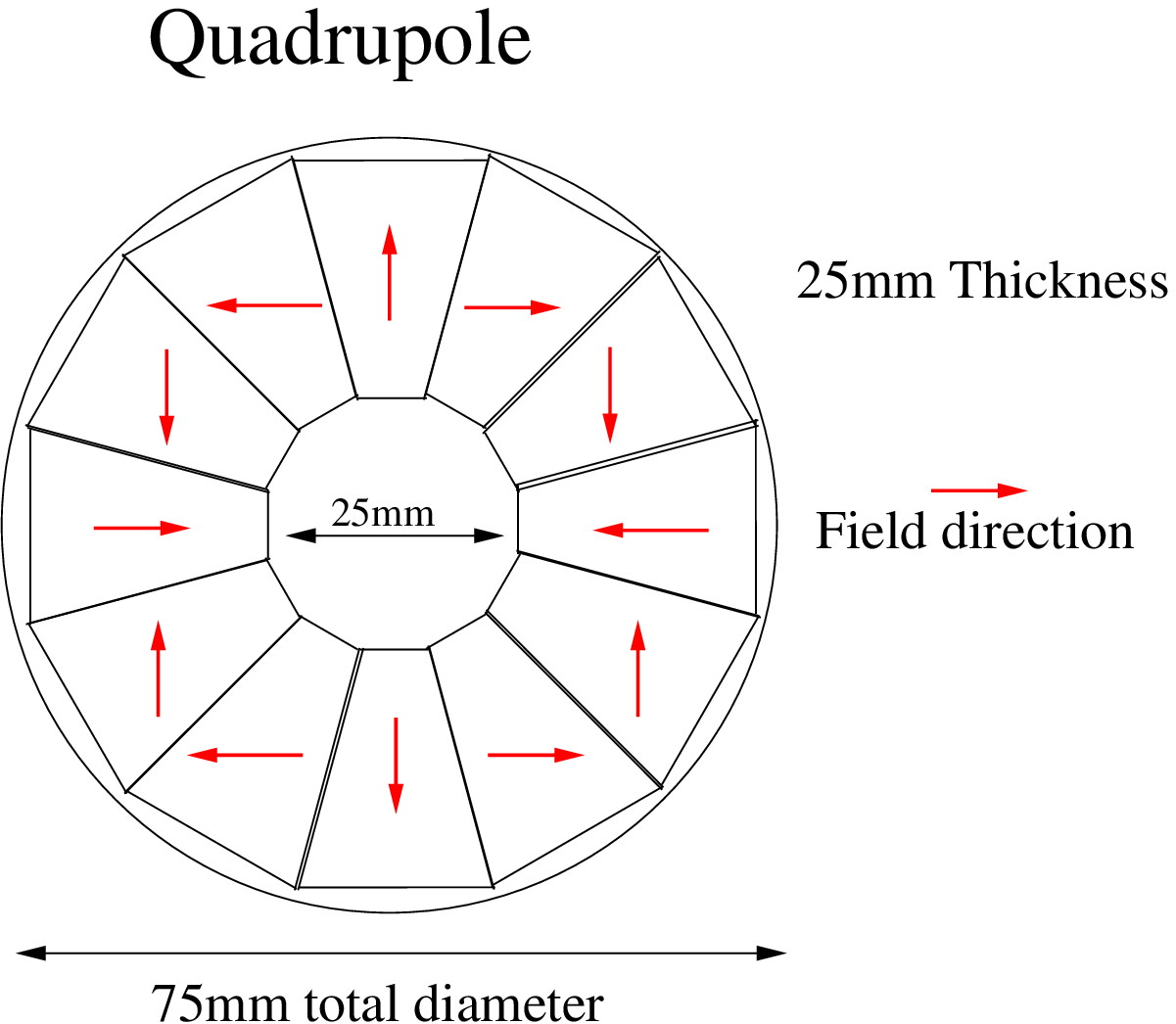}(b)\includegraphics[width=0.3\paperwidth]{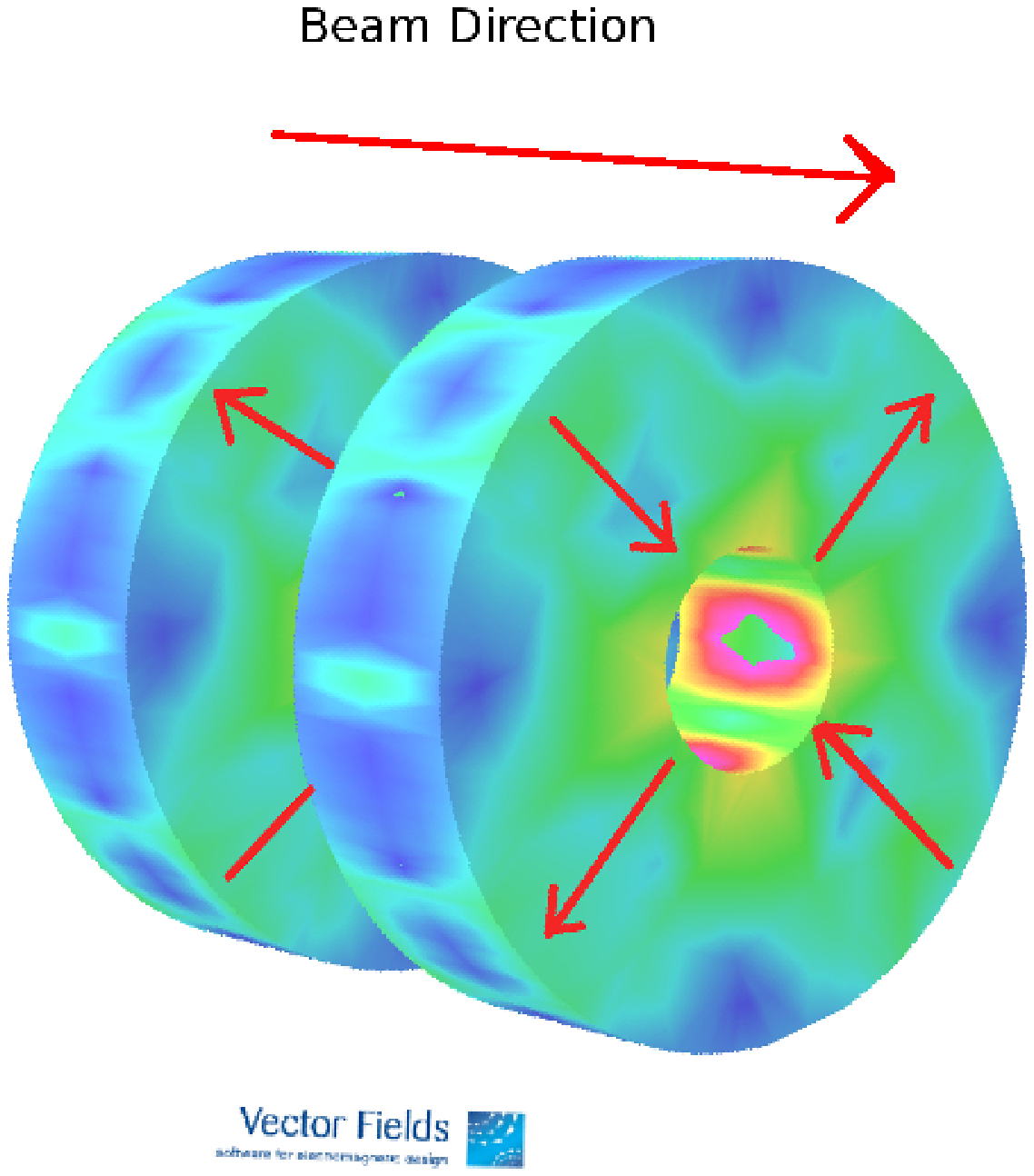}
\par\end{centering}

\raggedright{}\caption{The Halbach Cylinder Configuration(a) and the quadrupole focussing
pair(b).\label{fig:The-Halbach-Cylinder}}

\end{figure}

The quadrupoles employed in this experiment utilise the Halbach cylinder
design as shown on the left in Fig~\ref{fig:The-Halbach-Cylinder}.
Using this configuration has the effect of increasing the magnetic
flux density within the cylinder, and greatly reducing it outside\cite{halbach_cylinder}.
This has obvious benefits for use in compact portable systems. The
cylinder is constructed of 12 wedges of NdFeB alloy, where each wedge
has a remanent field strength of 1.36 Tesla, and in the configuration
shown in Fig~\ref{fig:The-Halbach-Cylinder}(a), this creates a typical
quadrupole field gradient of $\sim$110T/m. This number is mostly
governed by the inner diameter of the Halbach cylinder which in this
case is 25 mm. In general the quadrupoles were placed as close as
possible to each other, with a small gap necessary to facilitate their
separation since the attractive force between each Permanent Magnetic
Quadrupole (PMQ) is intense. In the pair configuration the second
PMQ was rotated at 90\textdegree{} with respect to the first, and
this creates a standard focussing quadrupole pair configuration\cite{PMQ_pair},
shown in Fig~\ref{fig:The-Halbach-Cylinder}(b). 

\begin{figure}[H]
\begin{centering}
\includegraphics[width=0.4\paperwidth]{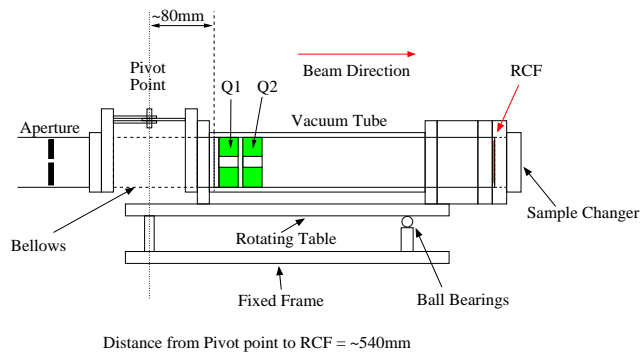}
\par\end{centering}

\caption{The overall set-up with magnets and radio-chromic film (RCF). \label{fig:The-overall-set-up}}

\end{figure}

Since this experiment describes the characterisation of this quadrupole
pair, a test-bed was constructed to allow the accurate placement of
the PMQs within a beam-line, and also to create fixed measurement
points for the detection system. Fig~\ref{fig:The-overall-set-up}
shows a vacuum-tight steel tube which is designed to house the PMQs
and various holders and spacers which regulate the distances and hold
the Halbach cylinder PMQ in place. This tube is attached to the Forschungzentrum
Dresden-Rossendorf Tandem Accelerator beam-line\cite{fzd_beamline}
which can produce protons beams of energy up to 9 MeV. 

The design incorporates a custom-built flange which places pieces
of radio-chromic film (RCF) at a specific distance downstream from
the PMQs (See Fig~\ref{fig:The-overall-set-up} for Sample Changer),
and the supporting structure for the vacuum tube allows for movement
of the tube with reference to a pivot point 8.5cm upstream from the
end of the tube. This has the effect of replicating the situation
where particles are transported into the PMQs with a specific angle
of incidence, and one of the main results measured shows the effect
of varying angles of incidence on the properties of transport and
focussing that the PMQs demonstrate.

\section{Method}

The experimental run at FZD was carried out using the Tandem accelerator
which produced proton energies of up to 9 MeV, at a typical beam current
of 1-10 picoamperes. This beam was tested, first without PMQs in position
to check the transport and alignment of the beam through the system,
then with one PMQ only, then with two PMQs arranged as a focussing
pair. The beam was passed through an aperture upstream of the pivot
point and tube, which regulated the beam spot-size. The PMQ pair was
typically placed at a distance of 25 mm apart which then defines the
strongest part of the field created by each PMQ as being 50 mm apart.
This reduced the extent to which the fields from a pair of PMQs interacted
and made them far easier to handle.

\subsection{Focal point analysis}

For the first part of the run a 1mm beam aperture was used, however
alignment issues forced the adoption of a 2mm aperture for later runs.
Once a suitable centre-point position for the moveable table supporting
the system was established, the aim was to fire the proton beam through
the PMQs in unchanging conditions, varying only the angle of incidence,
achieved by moving the support table a set distance from the centre-position,
in intervals corresponding to 1 degree. From this process, the effect
of the PMQs on incident 9 MeV proton beams with 2mm diameter and varying
angle of incidence could be recorded on a single piece of RCF. To
analyse the effect of the PMQ pair on the pencil beam produced by
the accelerator, the configuration in Fig~\ref{fig:The-overall-set-up}
was used and the incident angle was varied between 0-4 degrees, producing
an array of spots on the RCF which are discussed later.

\subsection{Definition of magnet focussing power and focal length}

For the case of a charged particle travelling with speed $v$ perpendicular
to a magnetic field $B$, the force experienced by the particle is
$F=qvB$. The resulting circular trajectory has a radius $\rho=\frac{mv}{qB}$,
varying proportionally with the particle's charge, momentum and the
strength of the external magnetic field. For a proton charge of $q=+1$,
$B\rho=mv$. The field gradient is defined as $K=\frac{dB_{y}}{dx}(Tm^{-1})$,
where the x-y plane is in the plane of the $B$-field. A quantity
taking account of the incident particle momentum and charge is the
normalised field gradient defined as $k=\frac{K}{B\rho}$, and this
is also known as the focussing strength of the quadrupole\cite{ParticleAccel}.
For a quadrupole where the region of the $B$-field has a length $L$,
the normalised field gradient applies over this distance to give a
focussing power of $\frac{1}{f}=\frac{KL}{B\rho}$, hence the focal
length is $f=\frac{B\rho}{KL}$, or in terms of the focussing strength,
$f=\frac{1}{kL}$. The PMQs shown in Fig~\ref{fig:The-Halbach-Cylinder}
have a measured field gradient of $100Tm^{-1}$, which for 9 MeV protons
gives a normalised field gradient of $k=268.5$, and a focal length
$f=\frac{1}{0.025*k}\,=\,0.15m$.

\begin{figure}[H]
\begin{centering}
\includegraphics[width=0.75\paperwidth]{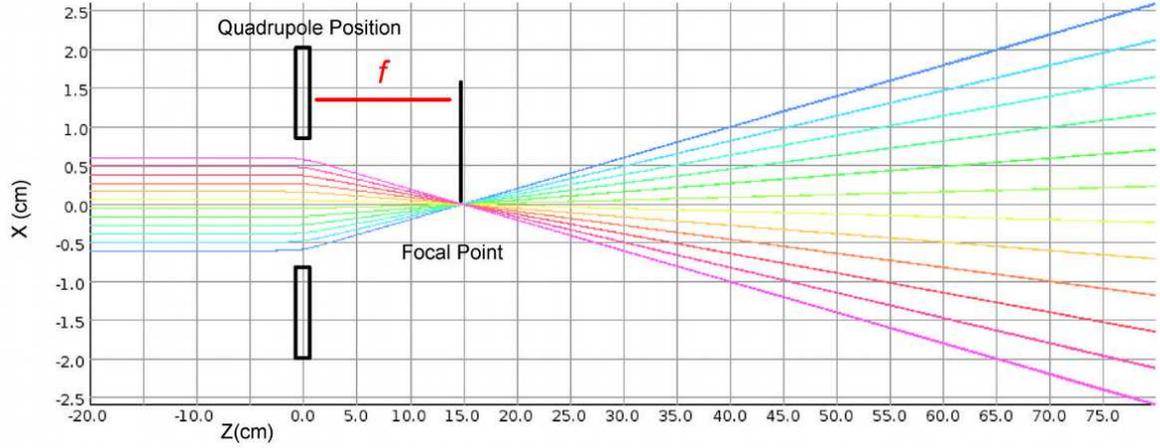}
\par\end{centering}

\centering{}\caption{TOSCA calculation of 9 MeV proton trajectory through a typical single
PMQ magnetic field.\label{fig:tosca_traj}}

\end{figure}

The focal length of 15 cm for this field gradient agrees closely with
proton trajectory calculations made in TOSCA\cite{vectorfields} (See
Fig~\ref{fig:tosca_traj}). This calculation shows only the focussing
effect in the x-plane, and there is an equal de-focussing effect in
the y-plane for a single PMQ. Using a combination of similar PMQs
in the FODO (focussing-zero-defocussing-zero) configuration gives
an overall focussing effect, as long as the separation distance between
the pair is less than the focal length of a single PMQ. If each PMQ
was infinitely thin, a single focal point would be the end product
of such a PMQ pair. In reality a finite dimension PMQ pair where the
first PMQ focusses in the X-direction will produce a focal line firstly
in X, then a similar orthogonal focal line in Y some distance downstream.

It is also possible to use a quadrupole triplet with a focussing-defocussing-focussing
effect (ABA configuration), which will constrain the diverging beam
in 1 dimension and tightly focus it in the other. There is a larger
amount of focussing overall in this arrangement than the traditional
PMQ pair, which makes this configuration worthy of further investigation. 

\subsection{Pencil beam ellipse analysis}

\begin{figure}[H]
\begin{centering}
\includegraphics[width=0.5\paperwidth]{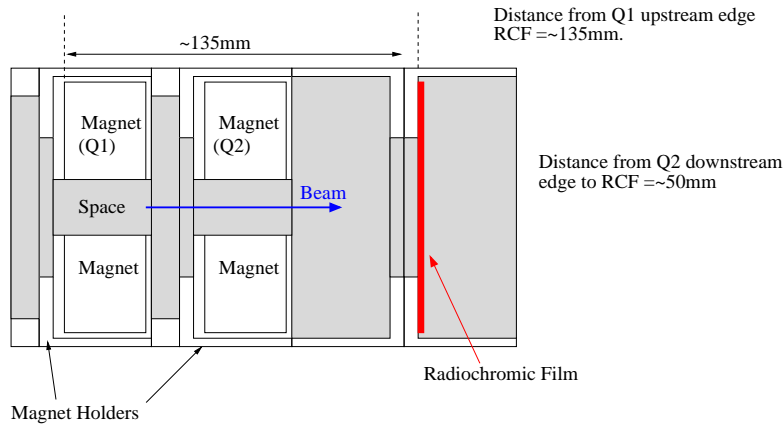}
\par\end{centering}

\caption{Shortened setup for analysing the typical ellipse produced by the
PMQ pair.\label{fig:ellipse-setup}}

\end{figure}

A useful way of characterising the PMQ pair is to look at the elliptical
distribution which defines the distortion of a circular acceptance
angle. This should be reproducible if the experimental set-up is accurate,
although a significant challenge arises from the fact that the support
table for the set-up can only move in one plane. One solution to this
is to rotate the PMQ pair through 45\textdegree{} and 90\textdegree{}
from the original position, and if the RCF is also rotated, this should
be consistent with the conditions producing the typical ellipse. A
modification of the setup used in Fig~\ref{fig:The-overall-set-up}
is used to ensure the placement of the PMQ pair is fixed with respect
to the RCF, seen in Fig~\ref{fig:ellipse-setup}. This arrangement
is more suitable for the purpose of measuring the elliptical acceptance
of the PMQ pair, as the RCF position and angle were fixed relative
to the PMQ orientation, which allows the results to be treated as
if the incident beam can travel at any angle deviating from the beam-line
rather than being limited to the horizontal plane, and greatly eases
the analysis of the RCF.

\subsection{Scattered divergent beam analysis}

\begin{figure}[H]
\begin{raggedright}
\includegraphics[width=0.4\paperwidth]{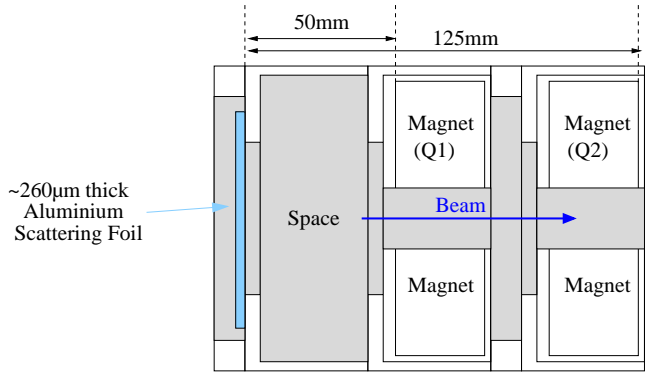}
\par\end{raggedright}

\caption{Shortened setup for analysing the effect of the PMQ on a divergent
beam produced by a scattering foil.\label{fig:divergent-setup}}

\end{figure}

Another method of characterising the quadrupole pair is to look at
the effect of the magnetic field on a divergent incident beam. The
method for this approach is to use a scattering foil upstream and
close to the first quadrupole, as shown in Fig~\ref{fig:divergent-setup}.
A comparison is then possible between the original scatter distribution,
with characteristic energy distribution and divergence determined
by the thickness and material of the radiator used as the scattering
foil, and the case where the quadrupole pair alter the trajectories
of the protons.

\subsection{Scattered divergent Beam with pepperpot beamlet analysis}

From limitations inherent in the experimental design, it follows that
better control over the incident angle is necessary for proper replication
of the results using this type of experimental set-up. A static method,
for generating several protons beams at once with varying incident
angles but equivalent experimental conditions passing into the test-bed
system, would give many experimental observables that would characterise
the PMQ pair and also greatly reduce the time spent on changing the
settings of the experiment. One way of meeting these goals is to use
a {}``pepperpot'', which is a foil or plate with many small holes
passing through it. The other component of this method is to use a
scattering foil as described in the Scattering Divergent beam analysis
section. If the holes are machined accurately, it is possible to deduce
the incident angles of any set of beams passing though the pepperpot
from a point-like source incident on the scattering foil if the distance
between the scattering foil and pepperpot is known. 

\begin{figure}[H]
\begin{centering}
\includegraphics[width=0.5\paperwidth]{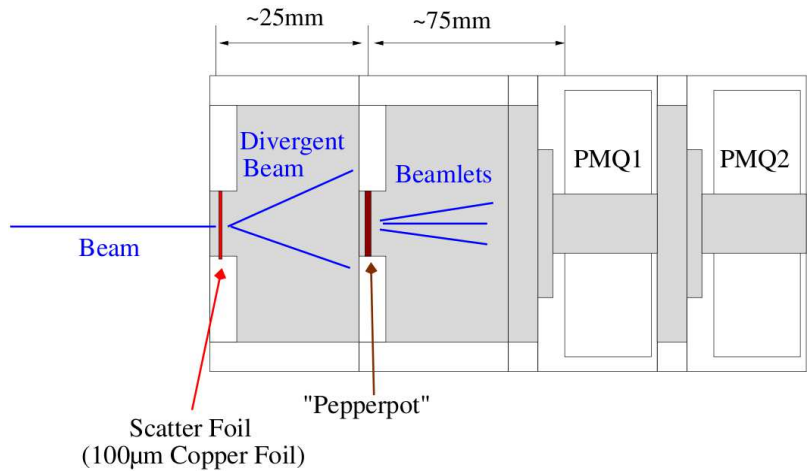}
\par\end{centering}

\caption{Simulated {}``Pepperpot'' set-up. An aluminium scattering foil was
simulated, with a thickness of 130\textmu{}m.\label{fig:Simulated-Pepperpot-set-up.}}

\end{figure}

Fig~\ref{fig:Simulated-Pepperpot-set-up.} shows the setup used at
FZD during the experiment. The pepperpot had a grid consisting of
121 holes of radius 50\textmu{}m and with a pitch of 1.25mm. This
pepperpot was created by using a micro-machining technique utilising
high-power femtosecond lasers to machine metal foil samples\cite{zhu_machining}. 

\begin{figure}[H]
\begin{centering}
\includegraphics[width=0.25\paperwidth]{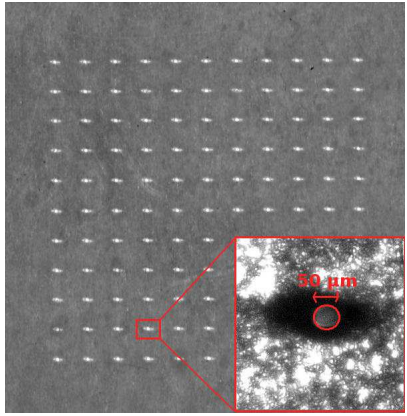}
\par\end{centering}

\caption{Example of Femtosecond Laser micro-machining produced at University
of Strathclyde with the TOPS laser system. The grid above consists
of 11x11 holes at a pitch of 0.5mm and 50\textmu{}m diameter.\label{fig:tops_micromachine}}

\end{figure}

Laser pulses of 800nm wavelength and 40 femtosecond pulse duration
are typically applied at a repetition rate of 1 kHz, with an energy
per pulse of 0.06 mJ, to drill holes via the laser ablation process
on a metal foil\cite{micromachine}. The machined holes are approximately
circular and have a diameter proportional to the laser focal point
diameter. The process is easily automated for the production of grids
and structures. Fig~\ref{fig:tops_micromachine} shows an example
of the quality and dimensions of holes produced in 250\textmu{}m thick
tungsten foil very similar to the one used in the experiment, using
the Femtosecond laser system of the TOPS facility at University of
Strathclyde.%

\begin{figure}[H]
\begin{centering}
\includegraphics[scale=0.6]{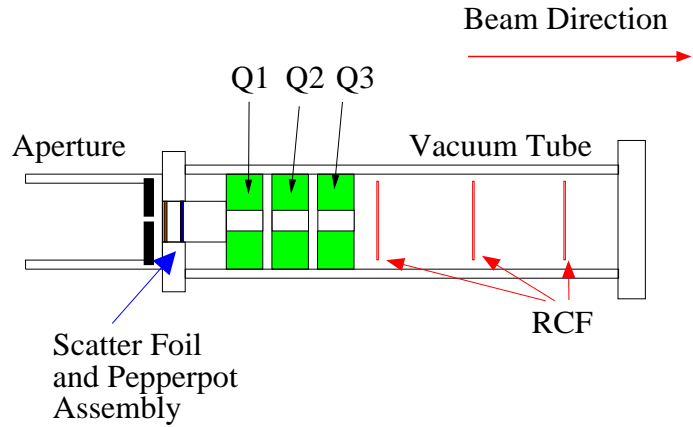}
\par\end{centering}

\caption{The modified experimental set-up used to implement the pepperpot.
RCF segments were placed at intervals along the length of the vacuum
tube to capture the evolving profile of the beamlets and their spatial
distribution.\label{fig:The-pepperpot_realsetup}}

\end{figure}

The set-up as described in Fig~\ref{fig:The-overall-set-up} must
be modified slightly as shown in Fig~\ref{fig:The-pepperpot_realsetup}
to accomodate the pepperpot, and to allow space for the placement
of RCF segments to record the beamlet positions and intensities. The
pivot and bellows are removed in this set-up since they were unneccesary.
A small 0.5mm radius aperture was used, and the scatter foil was placed
immediately downstream, with the rest of the geometry the same as
Fig~\ref{fig:Simulated-Pepperpot-set-up.}. A third quadrupole can
be optionally added to this set-up, and the results of the triplet
of quadrupoles placed in an A-B-A configuration (where B is orthogonal
to A around the z-axis), are discussed in the analysis section below.

\subsection{Selection of Scattering Foil and Pepperpot positions}

\begin{figure}[H]
\begin{centering}
\includegraphics[width=0.6\paperwidth]{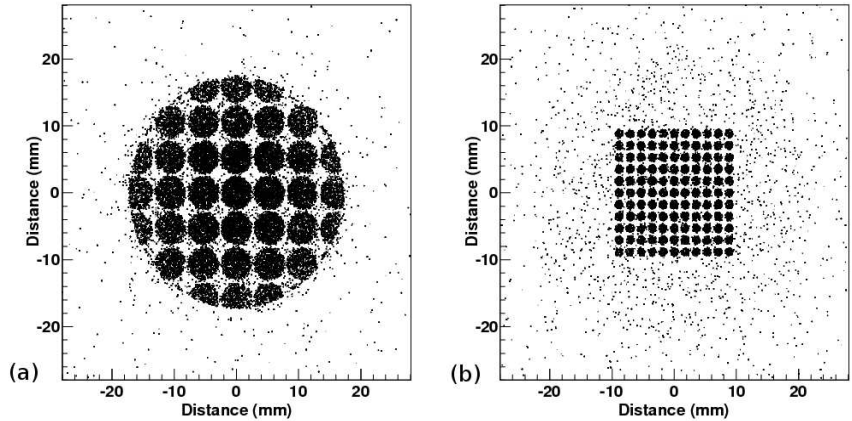}
\par\end{centering}

\caption{Figure~\ref{fig:beam_select}(a) demonstrates that a setup with the
pepperpot closer to the scatter foil produces clipping of the beamlets.
Figure~\ref{fig:beam_select}(b) shows that a setup with the pepperpot
further from the scatter foil allows them all to pass through.\label{fig:beam_select}}

\end{figure}

As demonstrated above, the focal length of an ion lens such as a quadrupole
magnet depends on the incident particle energy as well as the magnetic
field strength. Given the variety of possible usage of a system of
quadrupoles, it was decided to allow ample space between the scatter
foil, pepper pot and the quadrupoles. Trial and error allowed the
refinement of the placement of the pepperpot, since when placed closer
to the magnets, more of the beamlets pass through the system, and
the reverse happens when the pepperpot is placed closer to the scatter
foil. So the system can be optimised for the desired running conditions.
For the 9 MeV proton beam, which equates to a 6 MeV divergent beam
traversing the system after passing through the 100\textmu{}m copper
scatter foil shown in Fig~\ref{fig:Simulated-Pepperpot-set-up.},
placing the pepperpot 75mm downstream from the scatter foil allowed
all of the beamlets through, and placing the pepperpot 25mm from the
scatter foil greatly reduces the acceptance of the beamlets through
the system. Fig~\ref{fig:beam_select} shows a simulation of this
from GEANT4.

The angular resolution and acceptance of each beamlet changes as the
distance between the scatter foil and Pepperpot changes. For the configuration
in Fig~\ref{fig:beam_select}(a), each beamlet has an increased divergence
angle of 1.14$(\pm$0.11)\textdegree{}, while for the configuration
in Fig~\ref{fig:beam_select}(b), the increase in divergence angle
for each hole in the pepperpot is 0.38($\mbox{\ensuremath{\pm}0.04}$)\textdegree{}.
From this it is possible to estimate the acceptance of the magnet
system (limited by the inner bore of the PMQs when in the pair configuration)
when the source is at a distance of 10cm as being approximately 3\textdegree{}.

\section{Experimental Data and Comparison with simulation}

\begin{figure}[H]
\begin{centering}
\includegraphics[width=0.75\paperwidth]{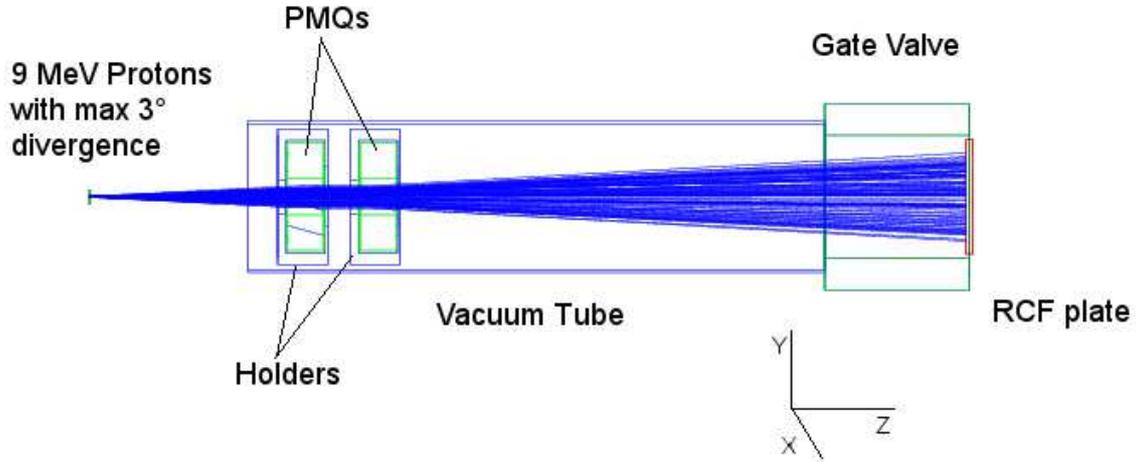}
\par\end{centering}

\caption{The GEANT4 simulation geometry. Dimensions exactly corresponding to
the experimental setup are shown above, with a 9 MeV proton beam with
3\textdegree{} opening angle displayed passing through the system
with the simulated magnetic field present. The relative orientation
of the system with respect to X,Y and Z axes in the simulation is
displayed. \label{fig:GEANT4-simulation-geometry.}}

\end{figure}

The results obtained are directly visible blue spots on RCF, of type
GAFCHROMIC HD-810. The intensity of the blue colour is proportional
to the radiation dose received, and for the scattering results below
the RCF was scanned to get an estimate of the radiation intensity.
The comparison with simulation is done using the CERN physics simulation
package GEANT4\cite{geant4}. A simulated field-map of the Halbach
cylinder design has been produced using TOSCA\cite{vectorfields},
a magneto-static scalar-field problem-solving package. This field-map
contains the magnetic field components at each point in space throughout
the design, even outside the magnets themselves, and GEANT4 can then
track charged particles through the field, using well-known stepping
algorithms such as the Runge-Kutta method\cite{butcher}. Comparison
with the RCF results is then achieved by using an equivalent geometry
to the real-world setup (see Fig~\ref{fig:GEANT4-simulation-geometry.}),
and analysing the predicted proton distribution at a given distance
downstream from the PMQs which corresponds with the actual positions
used for the RCF. 

\subsection{Focal point results and comparisons}

In Fig~\ref{fig:focussing-comparison} below, the results of firing
the accelerator proton beam at 9 MeV with a current of 1-10 picoamperes
for a duration of 15s per measurement are visible as blue ellipses.
As explained in the method we measured at a range of incident angles,
and this is shown in Figs~\ref{fig:focussing-comparison} and \ref{fig:Ellipse result},
where measurements from 0-4\textdegree{} are clearly shown. 

The simulation results are also shown in Fig~\ref{fig:focussing-comparison},
and the sizes and positions of the focal points obtained through the
PMQ pair, are comparable in size and position, suggesting that the
main source of experimental error here may be the positioning of the
rotating table.

\begin{figure}[H]
\begin{centering}
\includegraphics[width=0.65\paperwidth]{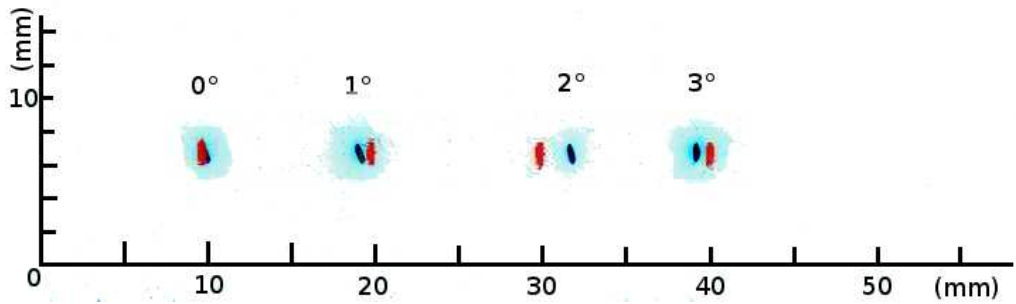}
\par\end{centering}

\caption{The RCF data is blue, and the GEANT4 prediction is red. The separation
between red and blue is a measure of angular accuracy. In this and
all other plots showing position and scale, the placement of the point
at 0\textdegree{} is arbitrary, so the positions given are relative
only.\label{fig:focussing-comparison}}

\end{figure}

\subsection{Pencil beam ellipse results and comparisons}

\begin{figure}[H]
\begin{centering}
(a)\includegraphics[width=0.3\paperwidth]{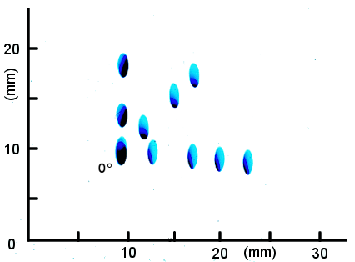}(b)\includegraphics[width=0.3\paperwidth]{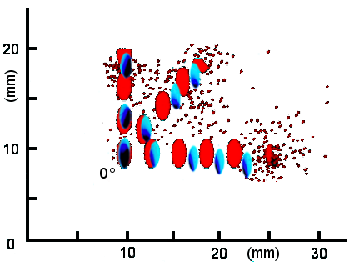}
\par\end{centering}

\caption{Result of using the setup described in Fig~\ref{fig:ellipse-setup}
and rotating the assembly through 2 steps of 45\textdegree{} each,
and varying the incident angle between 0\textdegree{} and 4\textdegree{}.
The experimental data is shown in (a) and the GEANT4 simulation is
overlaid for comparison in (b)\label{fig:Ellipse result}}

\end{figure}

For analysing the elliptical acceptance of the PMQ pair, and looking
at the consistency and quality of the focussing, measuring the resulting
position and shape of the pencil beam gives useful data, and in Fig~\ref{fig:Ellipse result}
the RCF shows the result of using the configuration discussed in the
method. The variation in distance between 1\textdegree{} and 2\textdegree{}
is present even when the PMQ-RCF assembly of Fig~\ref{fig:ellipse-setup}
is rotated, indicating a systematic error in the angular position.
There is clearly a much-increased angular acceptance in the x-plane
in Fig~\ref{fig:Ellipse result}, where protons with up to 4\textdegree{}
incident angle will still pass through the PMQ pair, and in the y-plane,
only 2\textdegree{} or lesser incident angle protons will pass through.
This is the expected behaviour of the PMQ pair, since charged particles
will typically be focussed in one plane first, then the orthogonal
plane second. Also, while focussing in one plane, a single PMQ disperses
in the orthogonal plane, and this effect dictates the acceptance of
the pair, along with position, particle $e/m$ and kinetic energy.

The results in Fig~\ref{fig:Ellipse result} show that the angular
acceptance is indeed elliptical, and the GEANT4 prediction agrees
with the general shape and size of each ellipse. The increased size
of each ellipse relative to the results in Fig~\ref{fig:focussing-comparison}
is due to the fact that the RCF is much closer to the PMQ pair than
in the original beamline setup, which is further evidence of the focussing
effect of the PMQ pair on a pencil beam.

GEANT4 predicts a more regular distribution, and two factors seem
to cause some distortion in the data. First, the angle of incidence
is underestimated in the experiment compared with the simulation where
these angles are known exactly. GEANT4 predicts almost 5\textdegree{}
angular acceptance in the x-plane which isn't corroborated by the
data. One straightforward explanation for this is the increased angular
step between 1\textdegree{} and 2\textdegree{} seen in the data due
to inaccuracies in obtaining the desired angle.

Second, there is also some evidence of an uneven distribution within
the beam spots themselves which could be due to: multiple scattering
effects arising from the beam passing through the aperture, or possibly
beam focussing and steering problems upstream from the test-bed location.
The consistency of this effect within the beam spots would seem to
indicate that this is a systematic effect characteristic of the experimental
set-up, and more work is required to investigate if this is reproducible
with similar beam parameters. The slight deviation in the x-plane
is due to the fact that the proton beam was not travelling exactly
through the centre of the magnets, and the effect on the position
is more pronounced at larger incident angles.

\subsection{Scattered divergent beam results and comparisons}

\begin{figure}[H]
\begin{centering}
(a)\includegraphics[width=0.35\paperwidth]{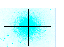}(c)\includegraphics[width=0.35\paperwidth]{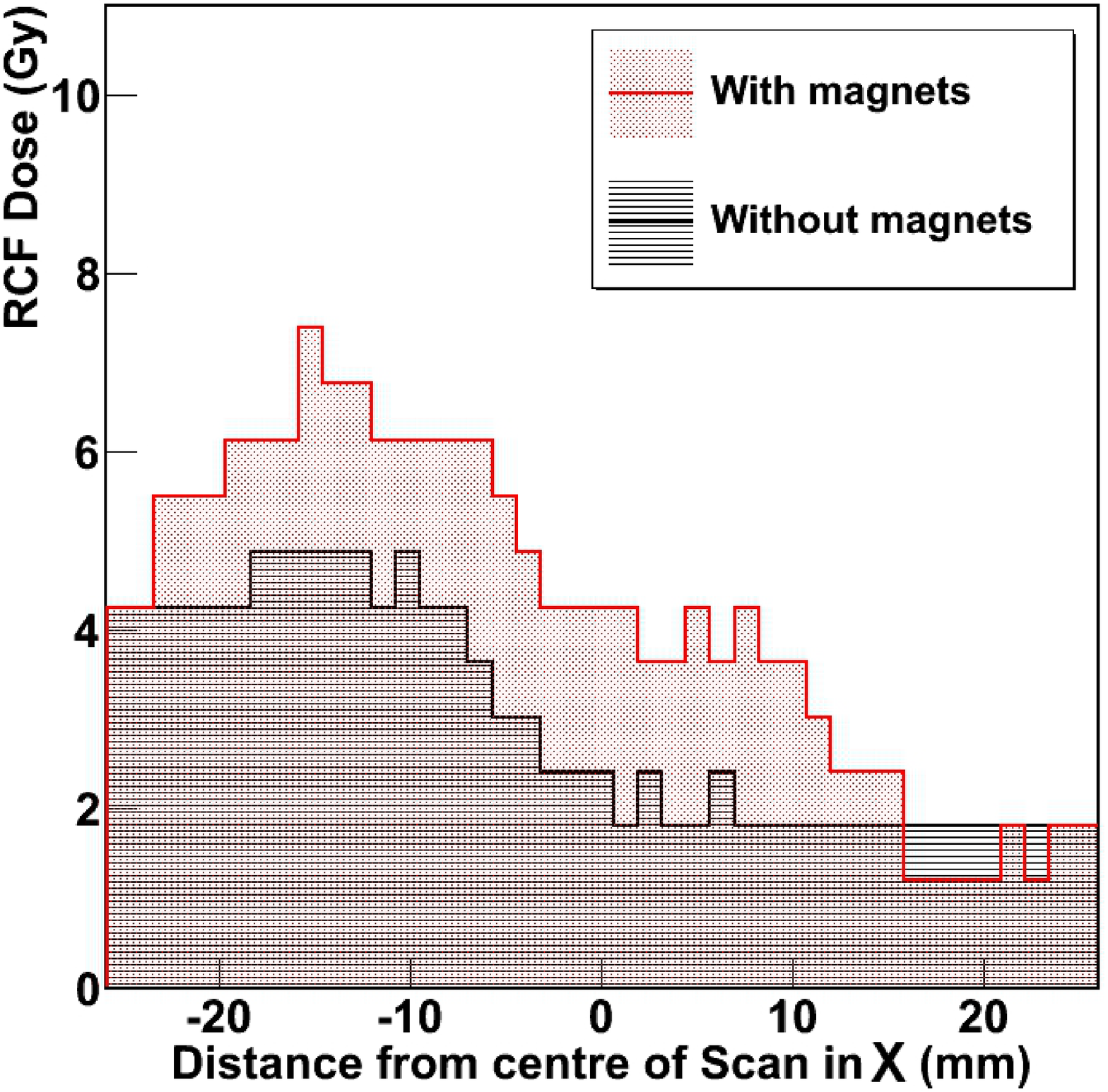}
\par\end{centering}

\begin{centering}
(b)\includegraphics[width=0.35\paperwidth]{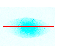}(d)\includegraphics[width=0.35\paperwidth]{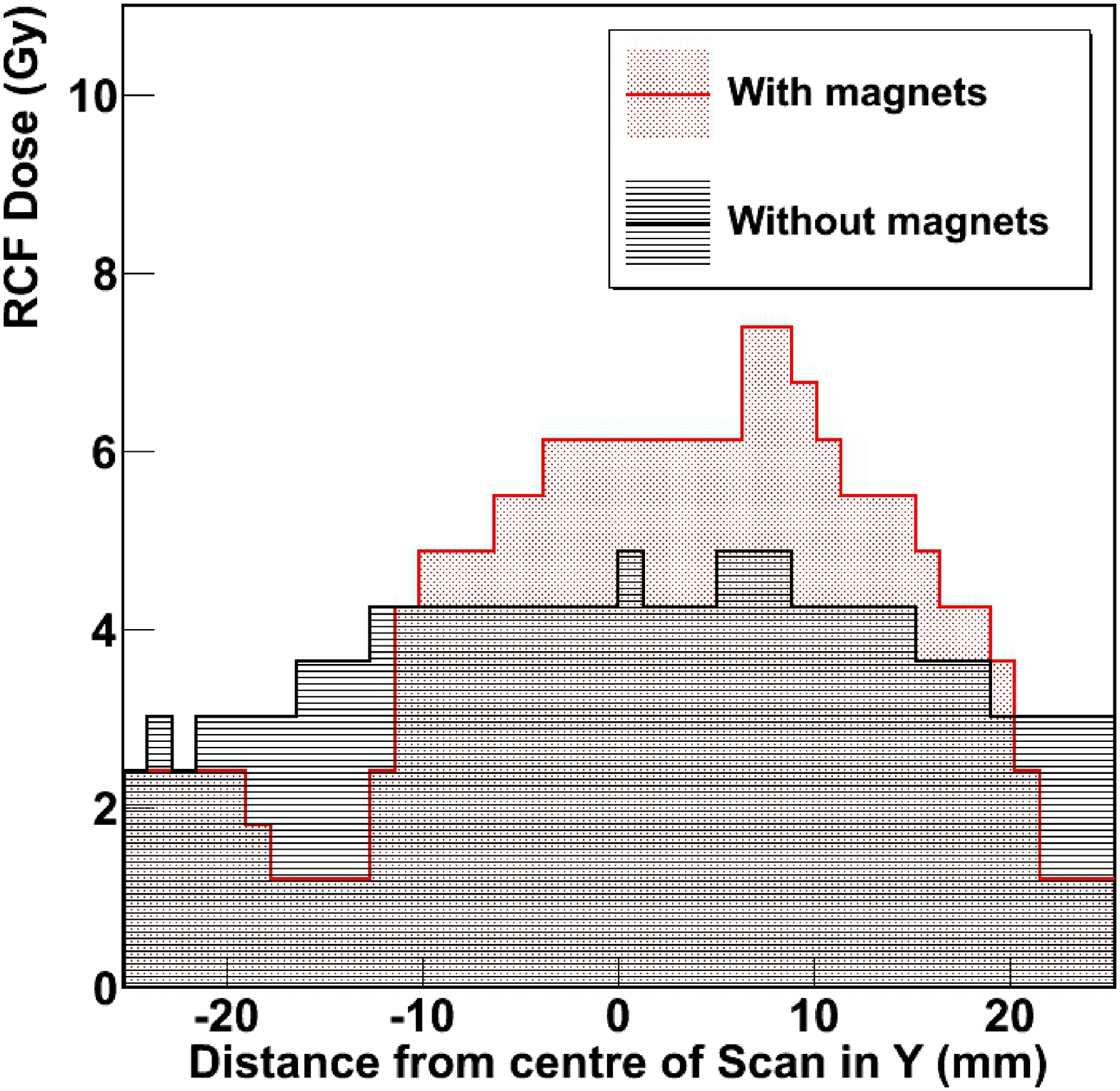}
\par\end{centering}

\caption{On the left, the RCF data showing the lines used for the profile analysis;
(a) unfocussed (b) focussed. On the right (c) the profile analysis
from both runs on the horizontal (X) plane, and (d) the vertical (Y)
plane. The distance offset from zero in (c) is due to limitations
in the area of RCF that the scanner could scan.\label{fig:scatter-results} }

\end{figure}

Using the configuration shown in Fig~\ref{fig:divergent-setup} the
results in Fig~\ref{fig:scatter-results} were obtained from a piece
of RCF placed at 400mm from the scattering foil. The mean proton energy
passing through the system when using a 260\textmu{}m aluminium scattering
foil is calculated from GEANT4 to be $\sim$4.75 MeV. The experimental
setup was not identical in both cases here since the unfocussed data
do not have the magnets present so the angular acceptance is not limited
by their inner diameter. However a comparison is still possible by
establishing the blue intensity through the central area of both plots.
This is done in both the X and Y planes by analysing the average blue
intensity over a user-selected line through the RCF to be analysed.

Although the acceptance difference between the two RCFs is obvious,
an analysis of the radiation dose of the central region of both RCFs
show a clear increase in dose in when the PMQ pair is present. Despite
the statistical errors due to low numbers of protons available for
analysis, the relative increase in yield from introducing the PMQs
into this divergent beam is estimated to be \textasciitilde{}25-35\%.

The data shown in Fig~\ref{fig:scatter-results} was analysed using
an adapted MATLAB routine\cite{martin} which uses a dose calibration
curve for the red, green and blue pixel values of a scanned RCF image.
This routine was modified to run in ROOT\cite{ROOT}, and was used
to generate a profile analysis of the radiation dose measured by the
RCF. The distance offsets seen in the profile analyses are due to
physical offsets from the centre of the RCF image introduced when
scanning the data. The scanner used is a Nikon Super COOLSCAN 9000
ED.

\begin{figure}[H]
\begin{centering}
(a)\includegraphics[width=0.35\paperwidth]{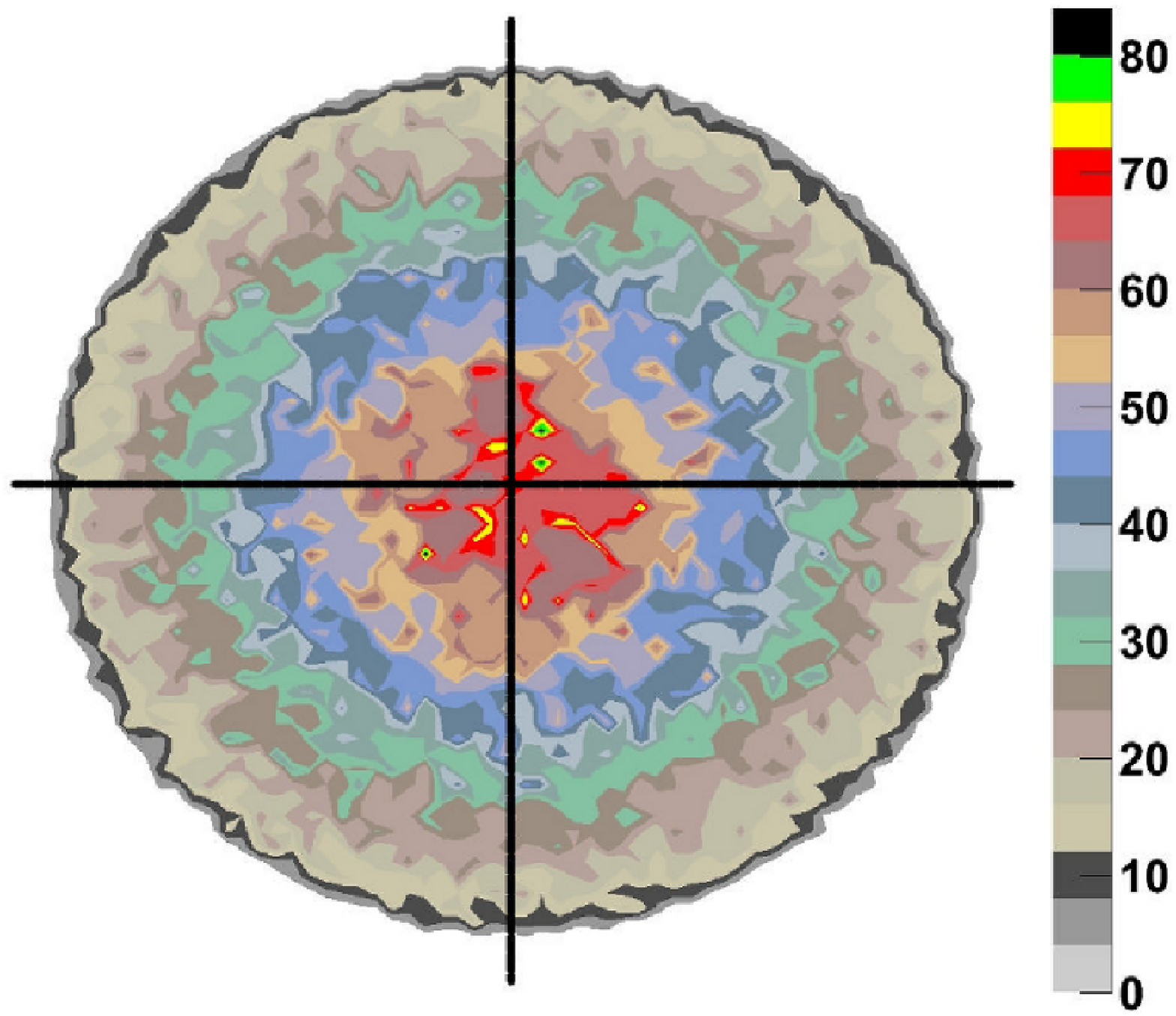}(c)\includegraphics[width=0.35\paperwidth]{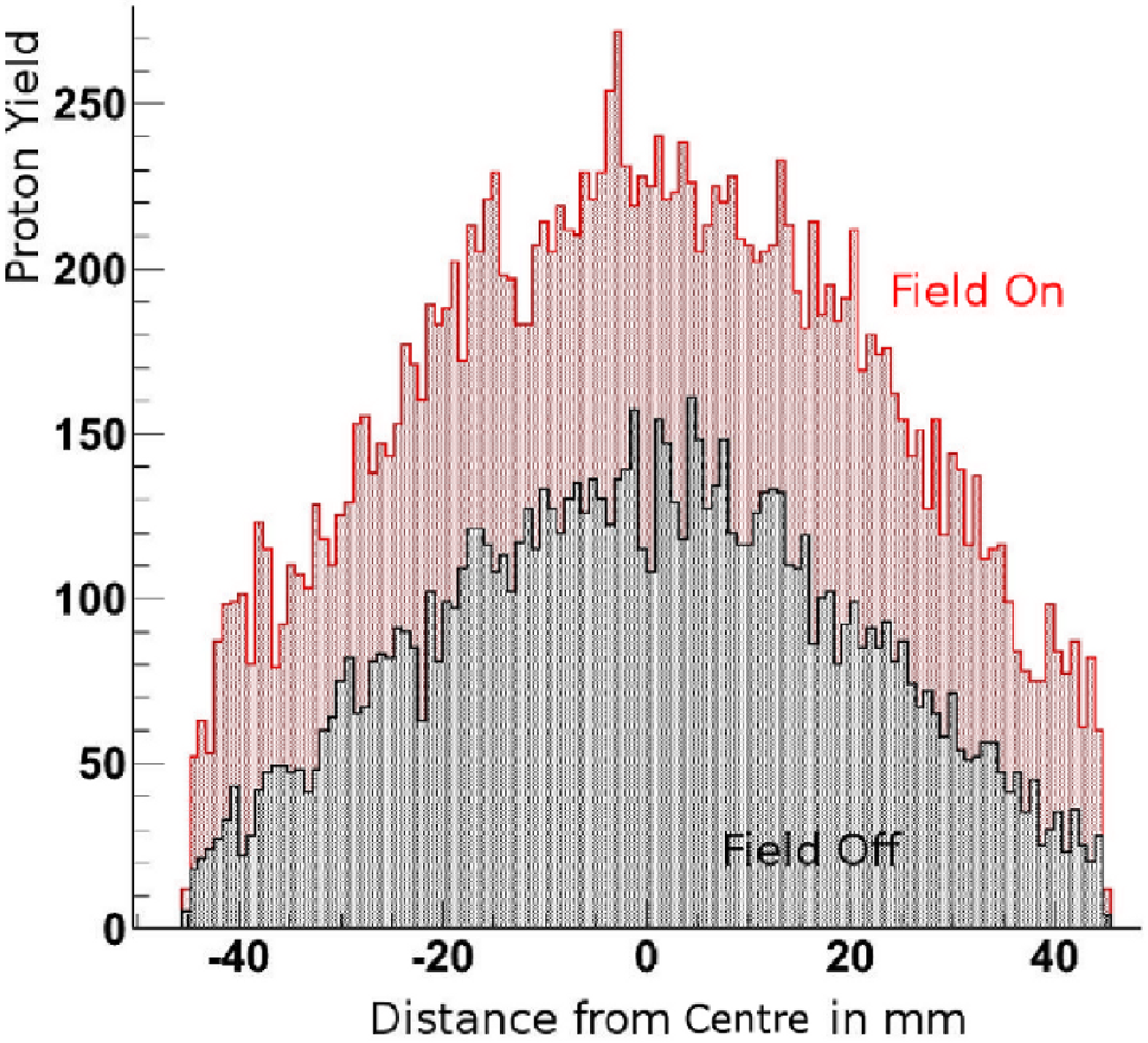}
\par\end{centering}

\begin{centering}
(b)\includegraphics[width=0.35\paperwidth]{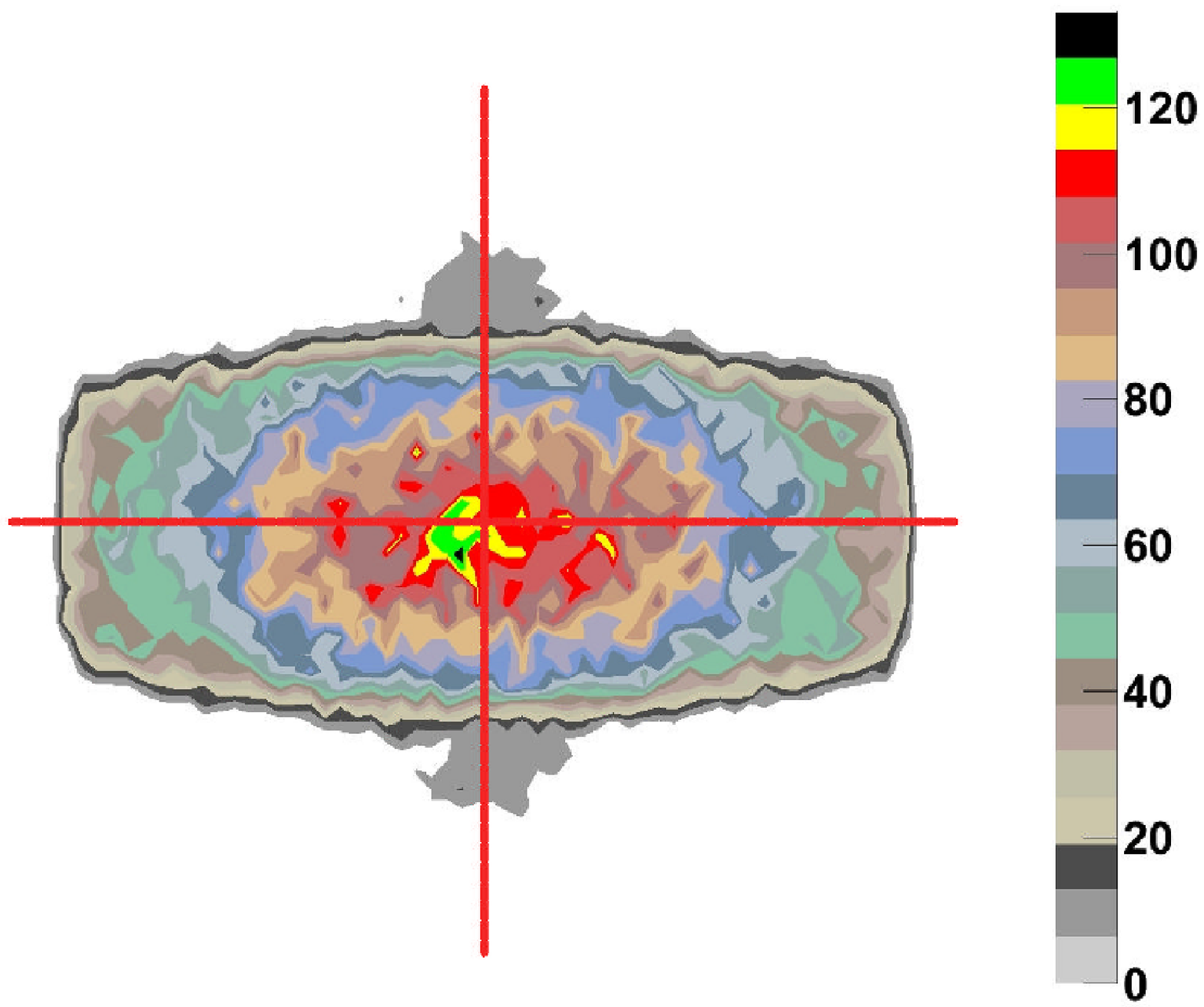}(d)\includegraphics[width=0.35\paperwidth]{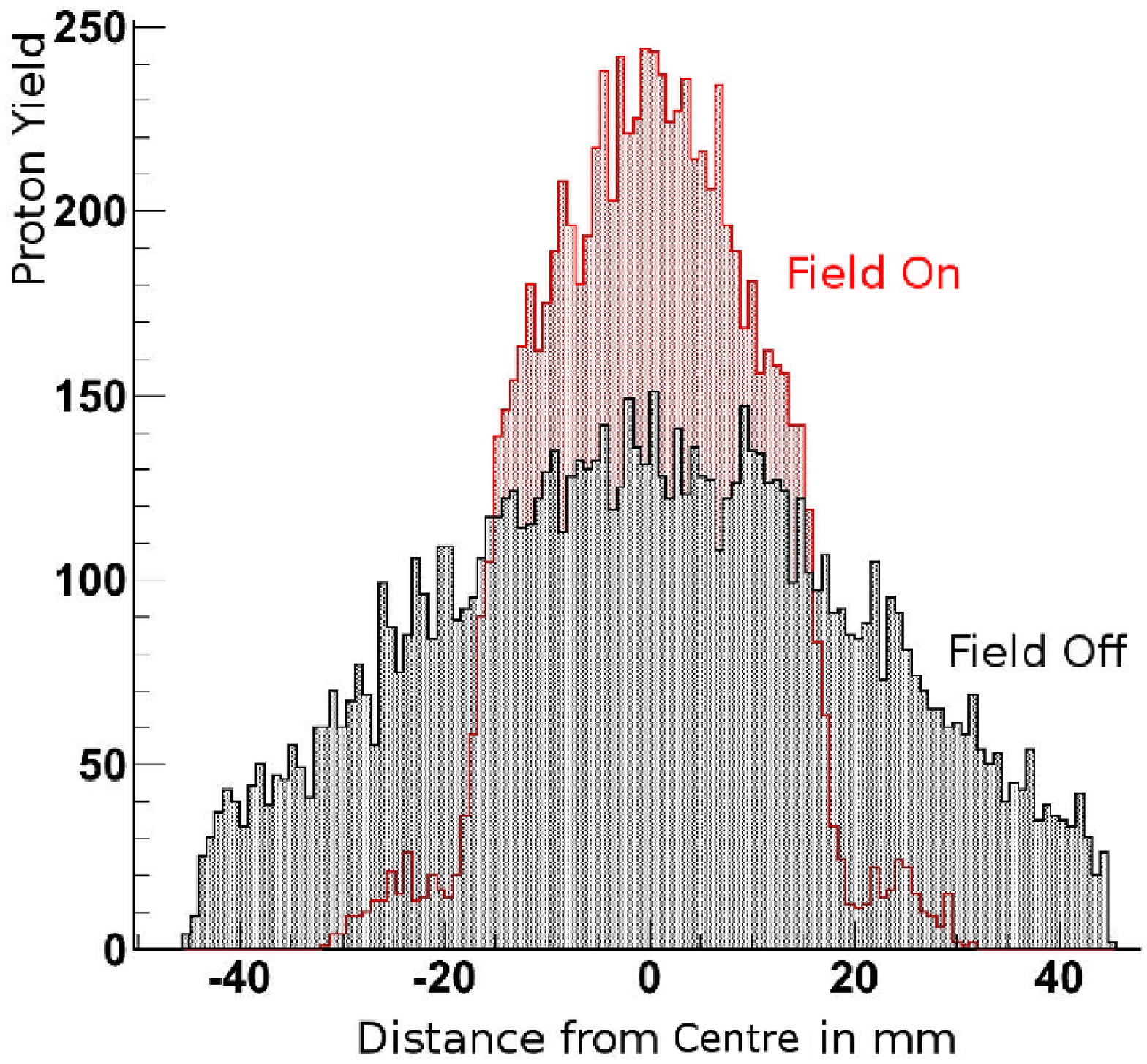}
\par\end{centering}

\caption{The GEANT4 simulation of the scattering beam setup. On the left, the
angular distribution at the RCF position (a) unfocussed (b) focussed.
On the right (c) a profile analysis in the horizontal (X) plane, and
(d) the vertical (Y) plane.\label{fig:g4_scatter_analysis}}

\end{figure}

A GEANT4 simulation of this setup was also carried out, using $10^{5}$
protons with 9 MeV initial energy passing through a scattering foil
of 260\textmu{}m thick aluminium . Fig~\ref{fig:g4_scatter_analysis}
shows the effects of the simulated PMQ pair on the scattered proton
beam, and there is a clear agreement in the shape and intensity of
the focussed ellipse with the experimental data shown in Fig~\ref{fig:scatter-results},
and this is also evident when the angular distribution of protons
is analysed at the RCF position in the simulation. A similar increase
in proton yield , with maximum enhancement at the beamline centre,
is seen in both X and Y planes, which indicates that the behaviour
of the simulated magnets is in agreement with the real PMQ pair.

\subsection{Pepperpot results and comparisons}

\begin{figure}[H]
\begin{centering}
(a)\includegraphics[width=0.3\paperwidth]{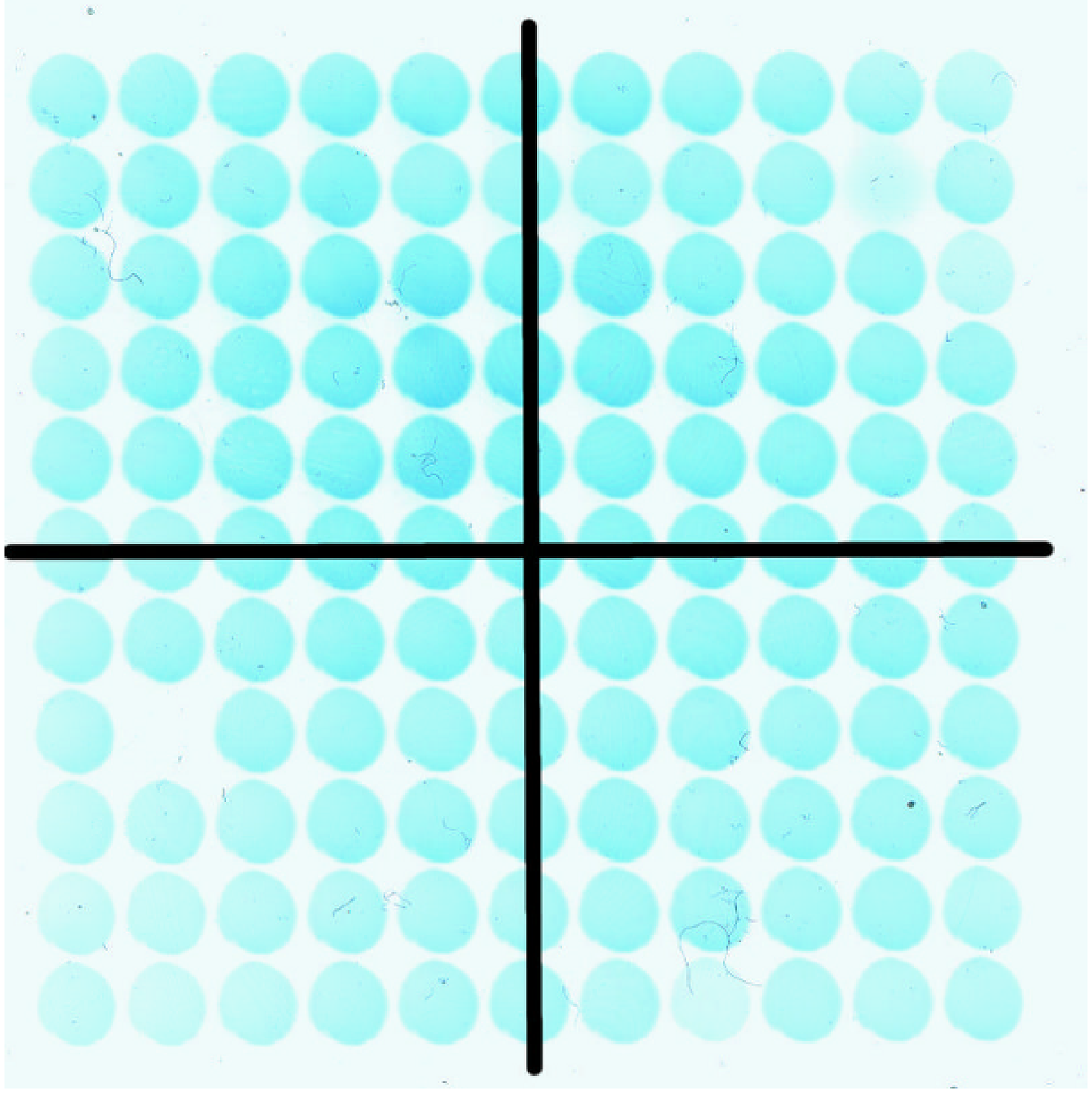}(b)\includegraphics[width=0.35\paperwidth]{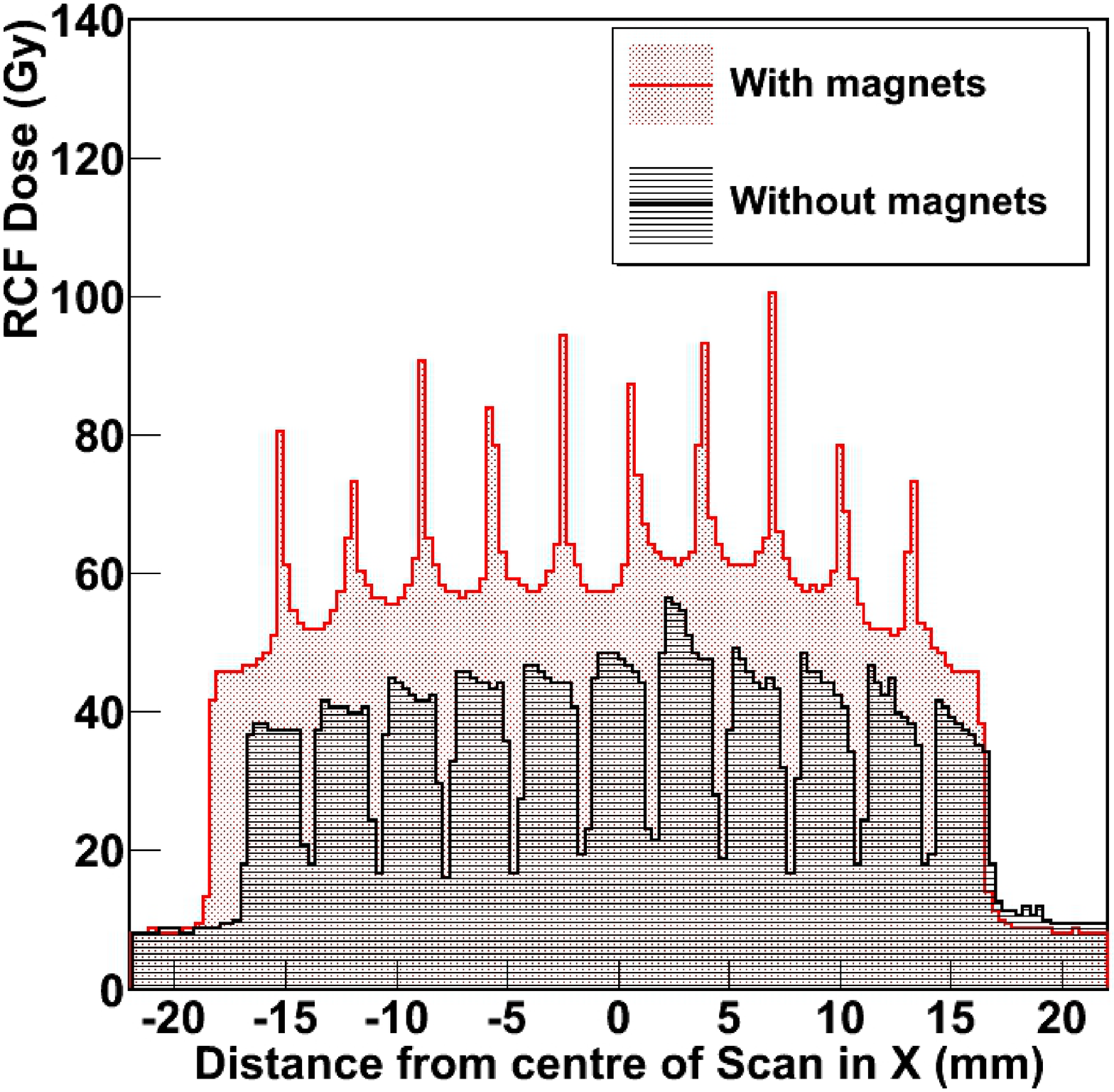}
\par\end{centering}

\begin{centering}
(c)\includegraphics[width=0.3\paperwidth]{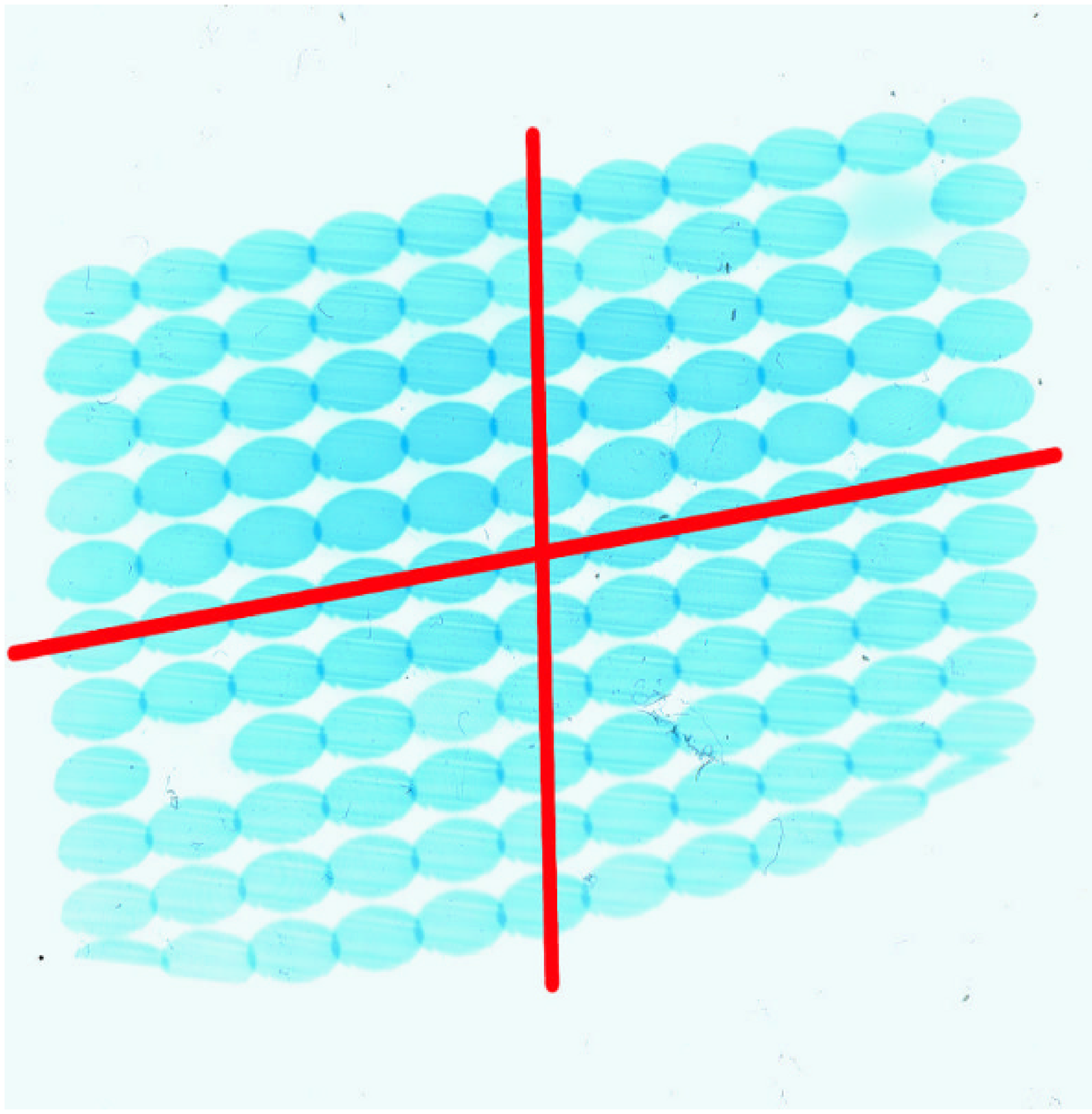}(d)\includegraphics[width=0.35\paperwidth]{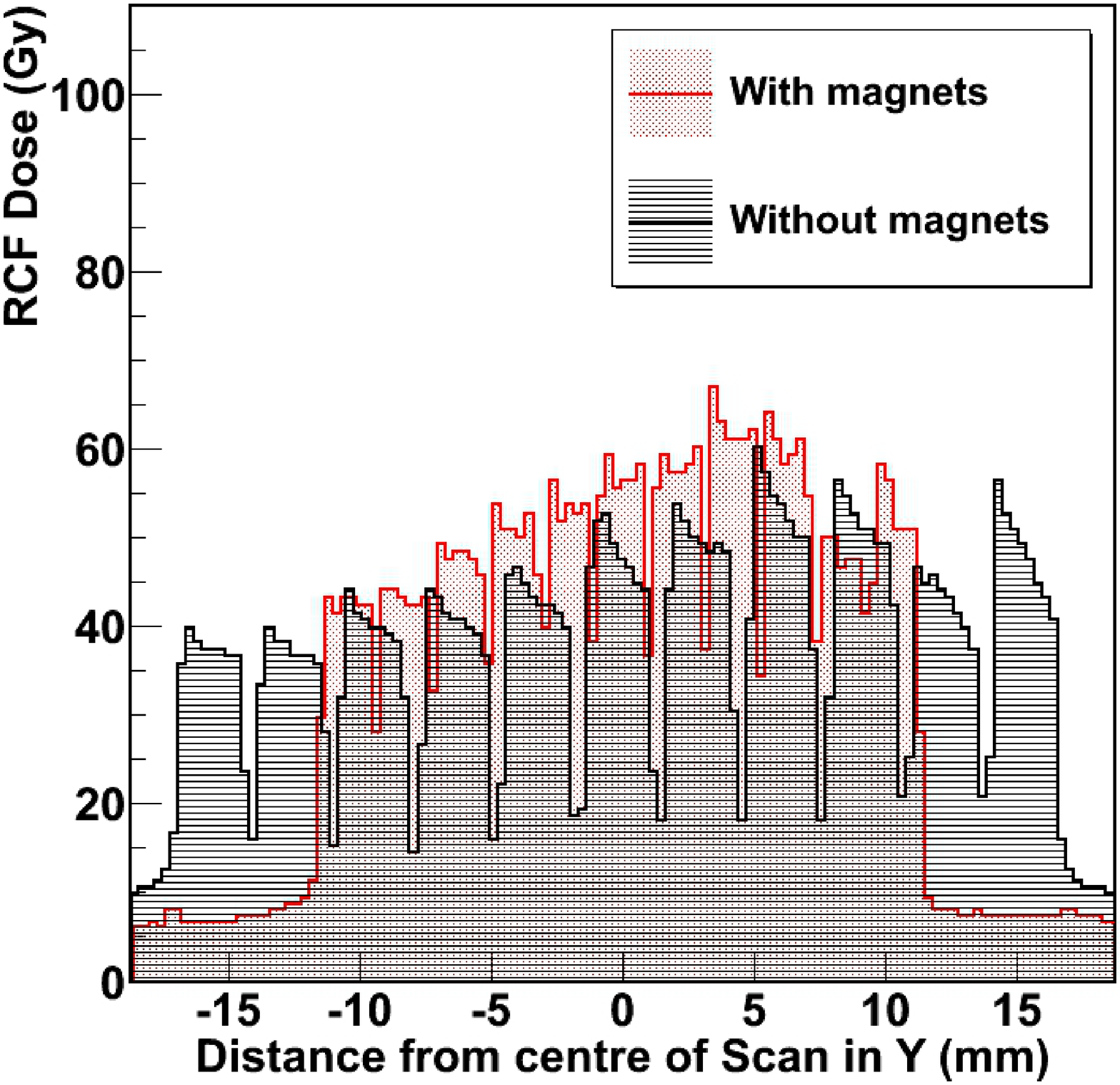}
\par\end{centering}

\caption{Pepperpot results on RCF 400mm downstream from scattering foil using
a 9 MeV tandem proton beam. On the left (a) shows the results with
no PMQs present, (c) shows the results with the PMQs present, and
on the right (b) shows a horizontal profile analysis comparing (a)
and (c), (d) shows the profile analysis performed in the vertical
direction. . The lines indicate the positions where the RCF was analysed.\label{fig:Pepperpot-results-real}}

\end{figure}

The original divergent proton beam produced by the scattering foil
reduces the tandem-provided 9 MeV down to an average energy of 5.97
\textpm{} 0.19 MeV. The beam is still very forward-peaked and has
an opening angle of $\sim$14 degrees, with both parameters measured
in GEANT4. A run using 7.5 MeV from the tandem was also used and this
provided a beam with an average proton energy of 3.79 \textpm{} 0.22
MeV and an opening angle of $\sim$ 16 degrees. The data produced
by the pepperpot and scattering foil on a single RCF sheet placed
at 400mm behind the scattering foil, with no magnets present, can
be seen in Fig~\ref{fig:Pepperpot-results-real}, where the gradient
in Fig~\ref{fig:Pepperpot-results-real}(c) is due to a mis-alignment
of the pepperpot about the beam axis, with an angle of $\sim$20 degrees.
Also clearly visible is an overlap between the beamlets in the X-direction
(See Fig~\ref{fig:Pepperpot-results-real}(b)), which almost doubles
the dose in these areas. This is not a direct focussing effect from
the PMQs, in fact it is due to the geometry of the pepperpot which
dramatically reduces the proton intensity at the RCF, so the effect
may be considered as a systematic error in the method.

\begin{figure}[H]
\begin{centering}
(a)\includegraphics[width=0.35\paperwidth]{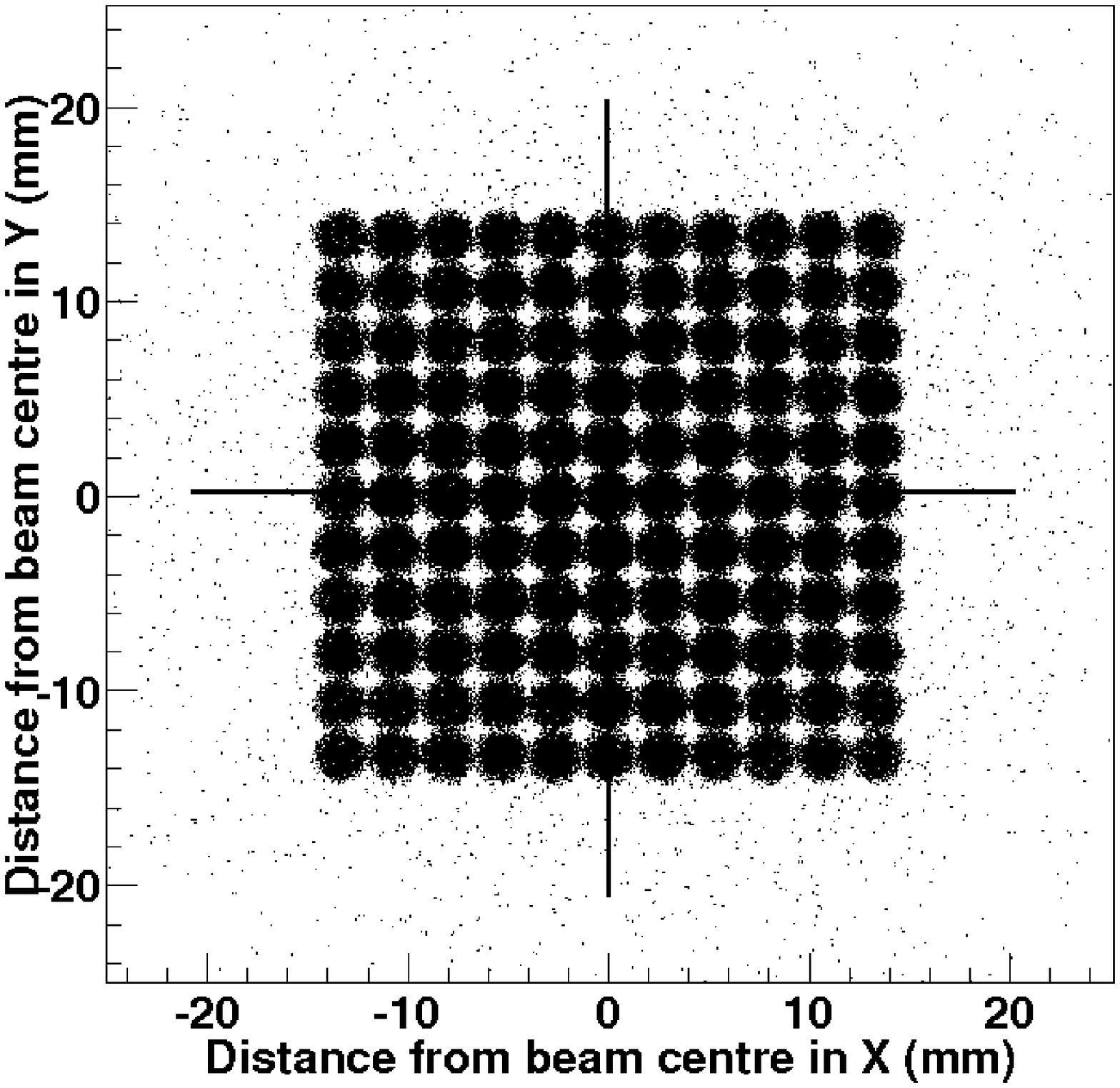}(b)\includegraphics[width=0.35\paperwidth]{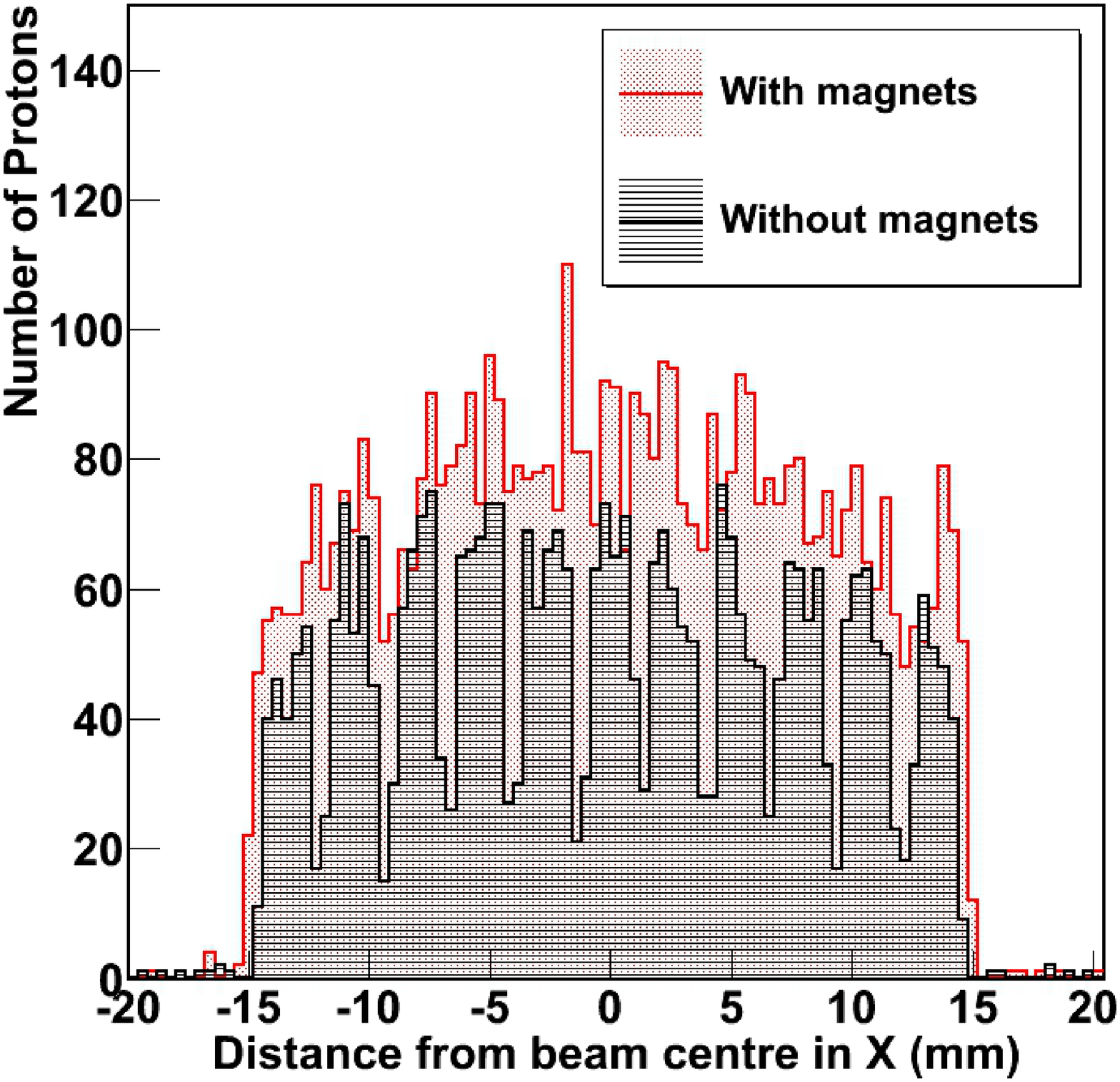}
\par\end{centering}

\begin{centering}
(c)\includegraphics[width=0.35\paperwidth]{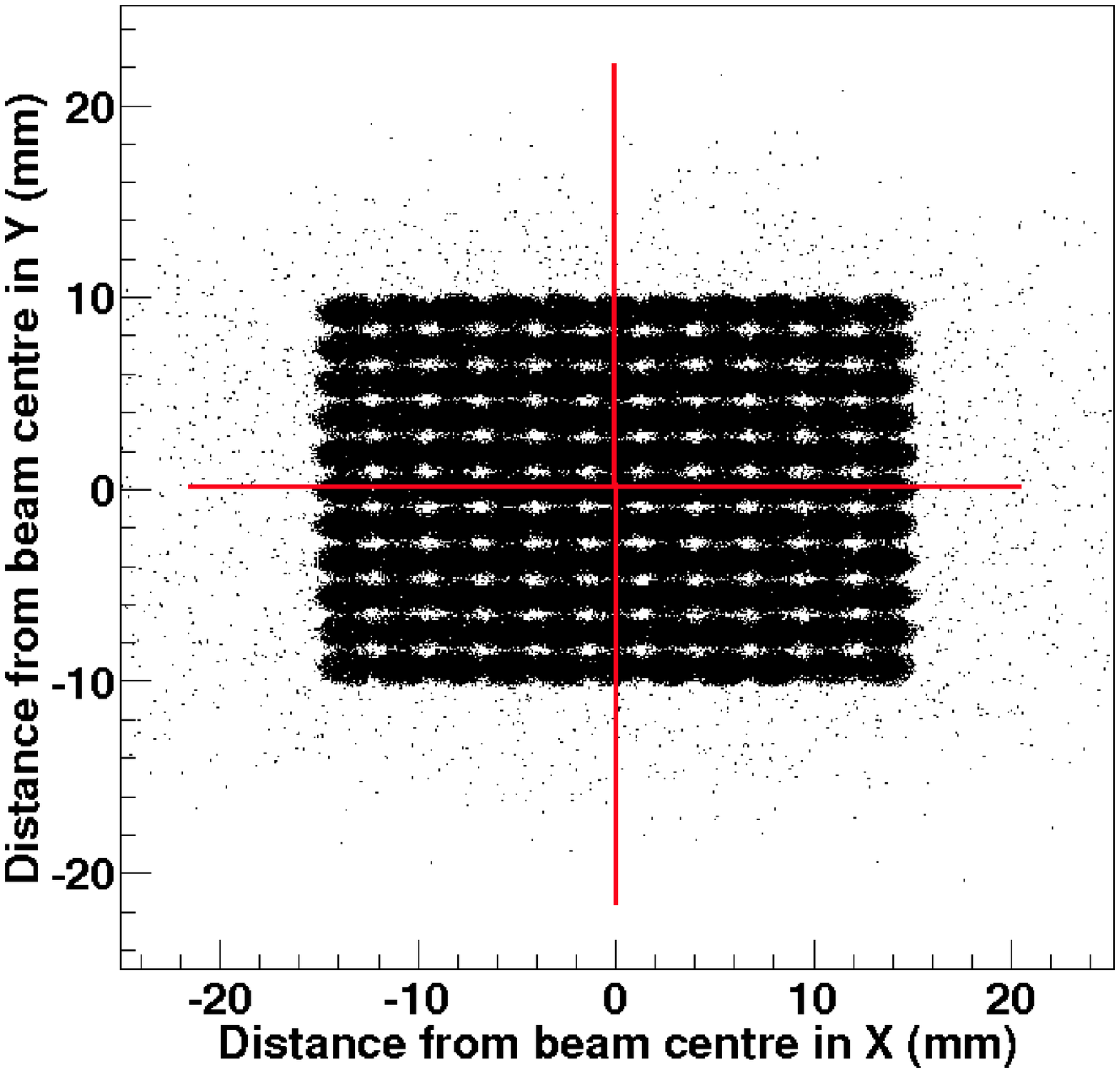}(d)\includegraphics[width=0.35\paperwidth]{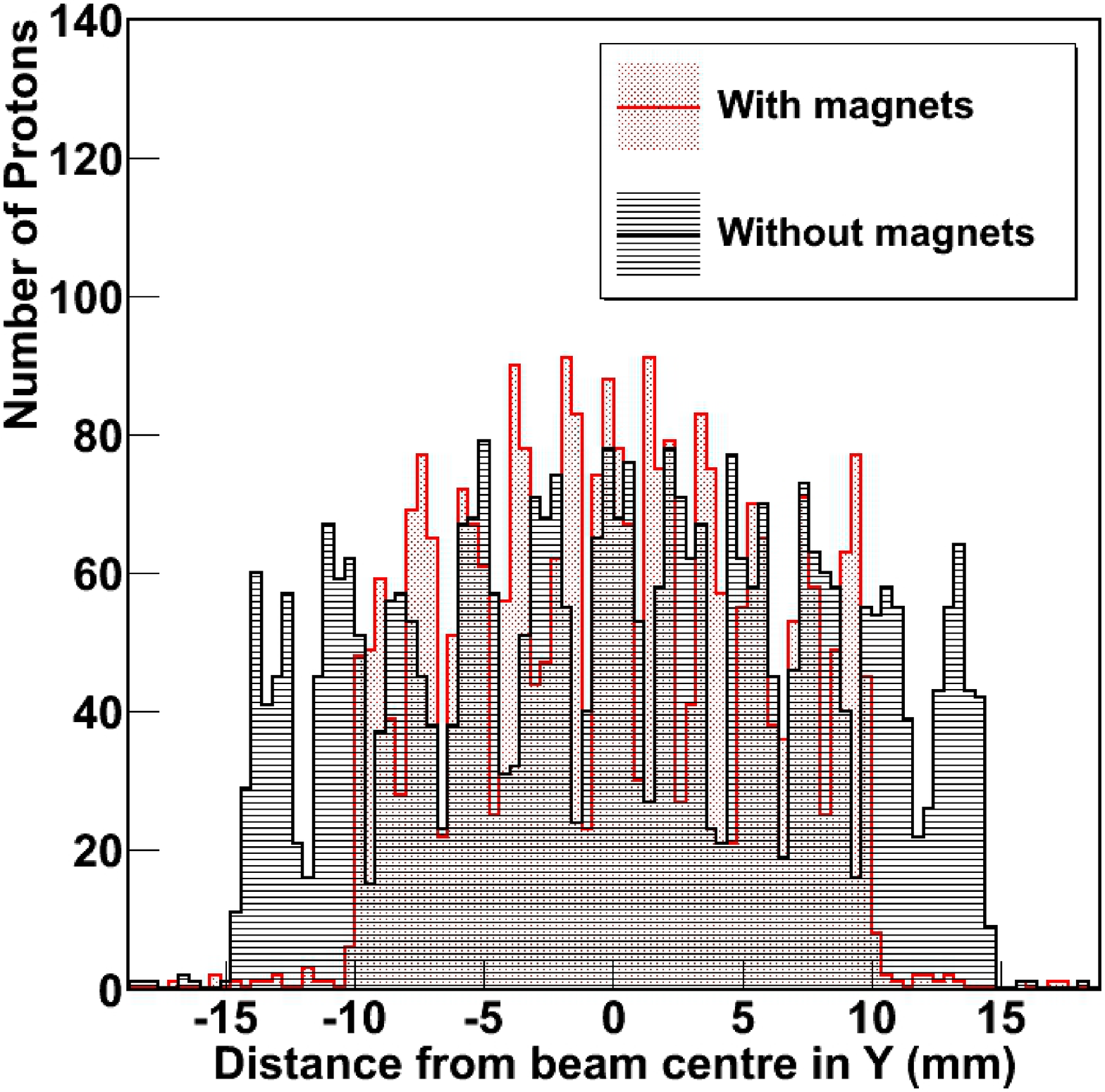}
\par\end{centering}

\caption{The GEANT4 simulation of the Pepperpot setup on RCF 400mm downstream
from scattering foil. On the left (a) shows the results with no PMQs
present, (c) shows the results with the PMQs present. On the right
(b) shows a horizontal profile analysis comparing (a) and (c), (d)
shows the same analysis performed in the vertical direction.The lines
indicate the positions where the RCF was analysed.\label{fig:Pepperpot-results-9MeV}}

\end{figure}

A GEANT4 simulation of $2\times10^{8}$ protons was carried out using
the same experimental parameters and the field-map produced by TOSCA.
The equivalent data and analysis are shown in Fig~\ref{fig:Pepperpot-results-9MeV}.
It is clear that the focussing and enhancement in intensity is less
here than in Fig~\ref{fig:scatter-results}, due to the higher proton
energy.

\subsection{Transport Properties for Diverging beam}

\begin{figure}[H]
\begin{centering}
\includegraphics[width=0.7\paperwidth]{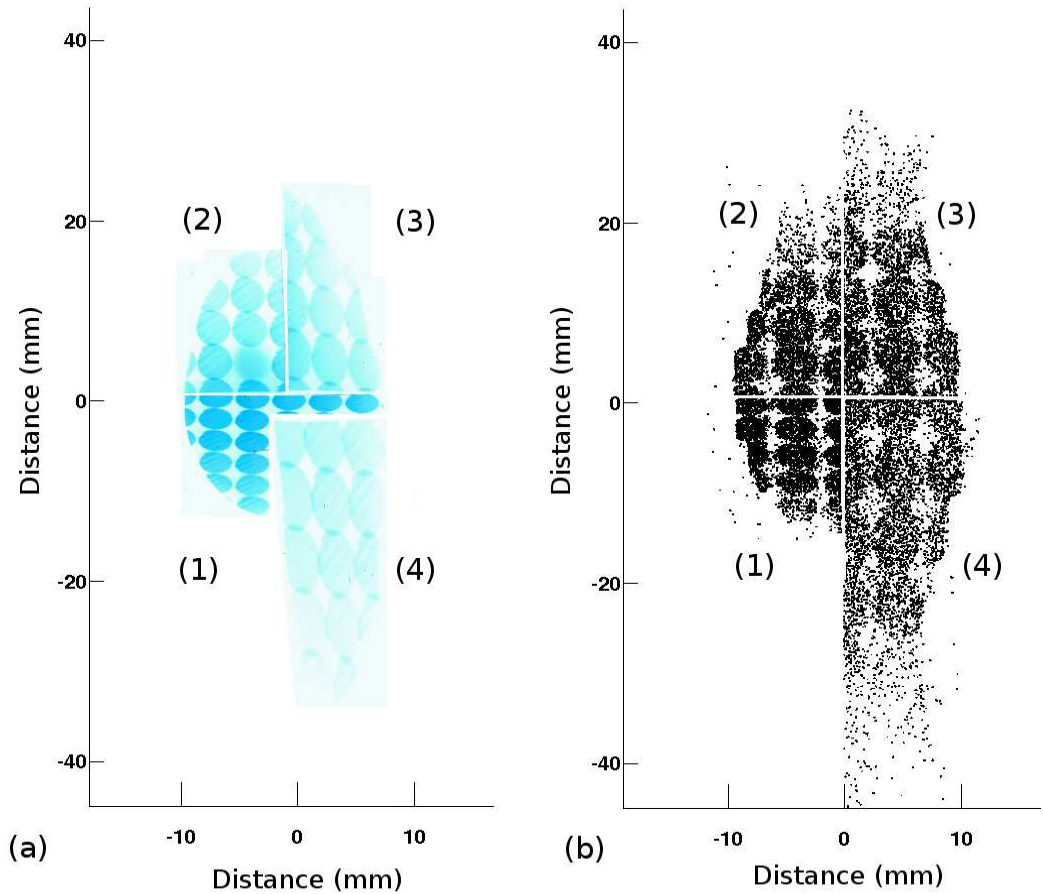}
\par\end{centering}

\caption{(a) Transport of the beam through 2 quadrupoles at 7.5 MeV using four
RCF segments placed at 200mm(1), 276mm(2), 352mm(3) and 442mm(4) from
the scatter foil. (b) Matching GEANT4 simulation run using 2x10$^{7}$
7.5 MeV protons and matching RCF positions.\label{fig:transport_quads}}

\end{figure}

As a measure of the transport qualities of the beam, a number of RCF
segments were used at known distances downstream of the PMQs to allow
study of the evolution of the beam distribution as a function of distance.
Firstly, a quadrupole pair was used with a tandem proton beam energy
of 7.5 MeV. Four segments of RCF were used in the vacuum tube to obtain
a profile of the beam downstream of the magnets, and the results of
this are shown in Fig~\ref{fig:transport_quads}(a). The simulation
results from GEANT4 are compared with the results in Fig~\ref{fig:transport_quads}(b),
and the main features of the data are reproduced. In terms of angular
distribution and acceptance angle of the incident beam, the simulations
match what was measured during the experiment.

\subsection{Quadrupole triplet focussing}

\begin{figure}[H]
\begin{centering}
(a)\includegraphics[width=0.5\paperwidth]{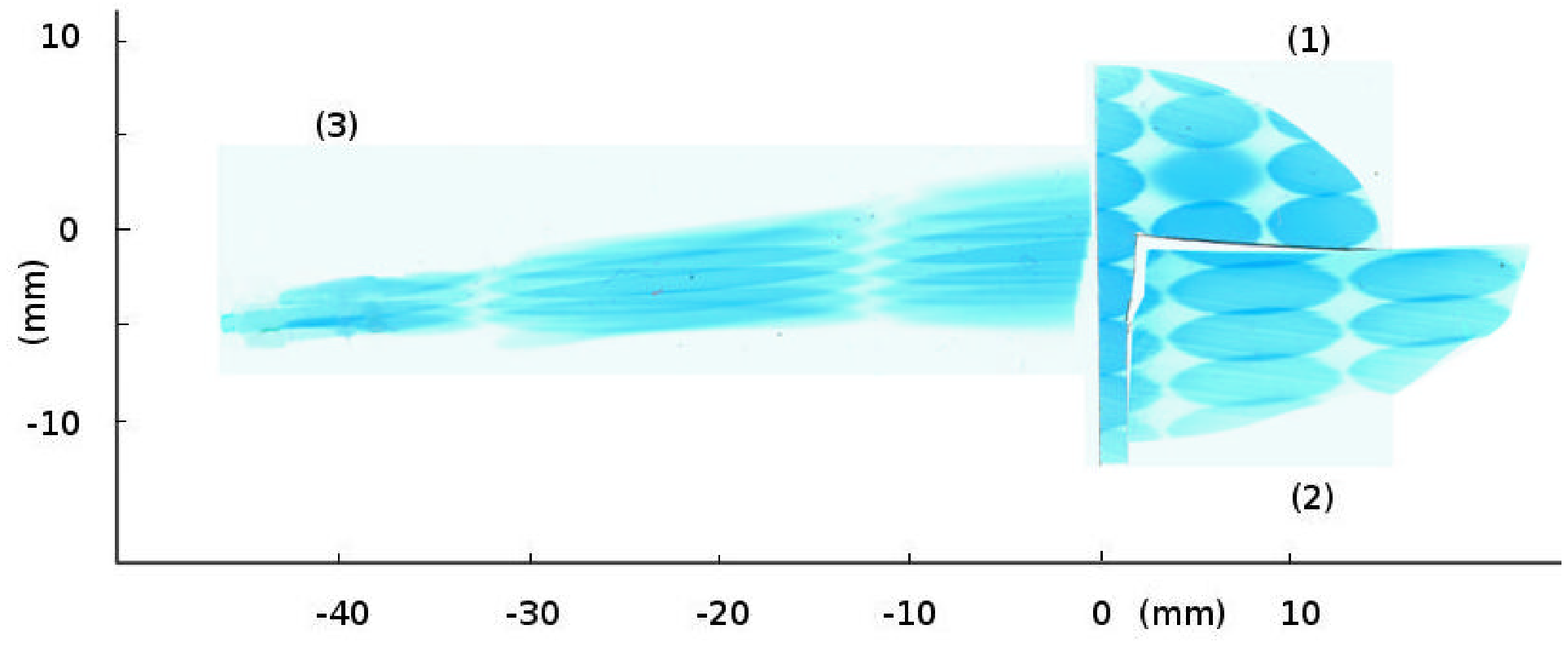}
\par\end{centering}

\begin{centering}
(b)\includegraphics[width=0.5\paperwidth]{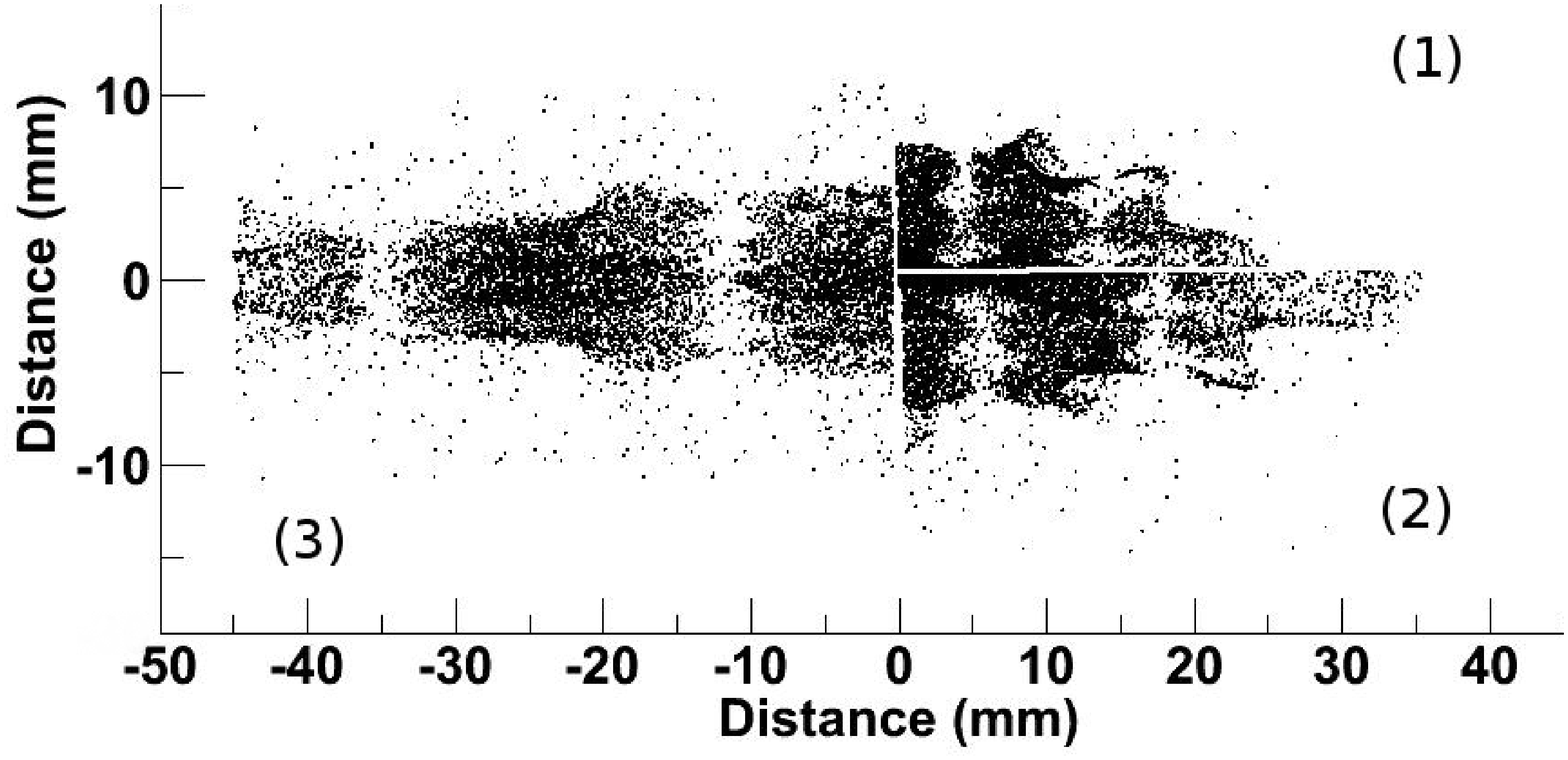}
\par\end{centering}

\caption{(a) Transport of the beam at 9 MeV using three RCF segments placed
at 296mm(1), 372mm(2) , and 620mm(3) from the scatter foil. (b) Matching
GEANT4 simulation run using 2x10$^{7}$ 9 MeV protons and matching
RCF positions.\label{fig:transport_triplet_9}}

\end{figure}

\begin{figure}[H]
\begin{centering}
(a)\includegraphics[width=0.5\paperwidth]{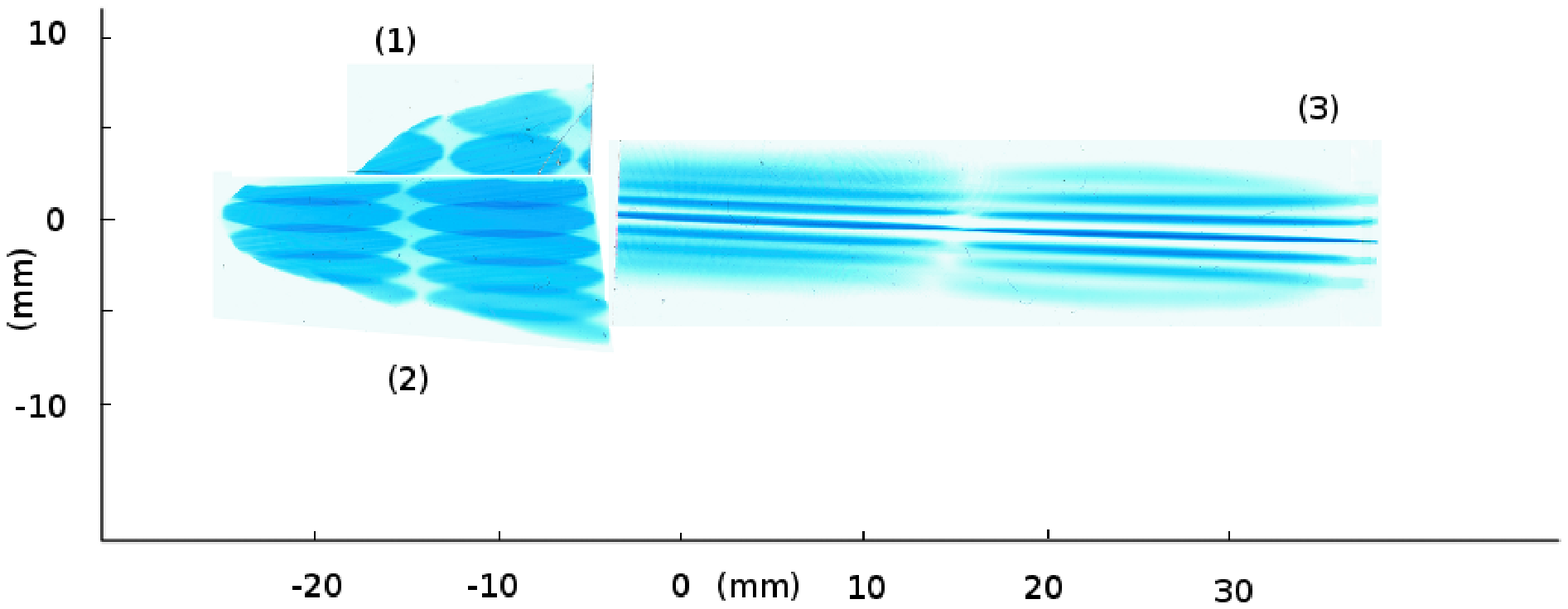}
\par\end{centering}

\begin{centering}
(b)\includegraphics[width=0.5\paperwidth]{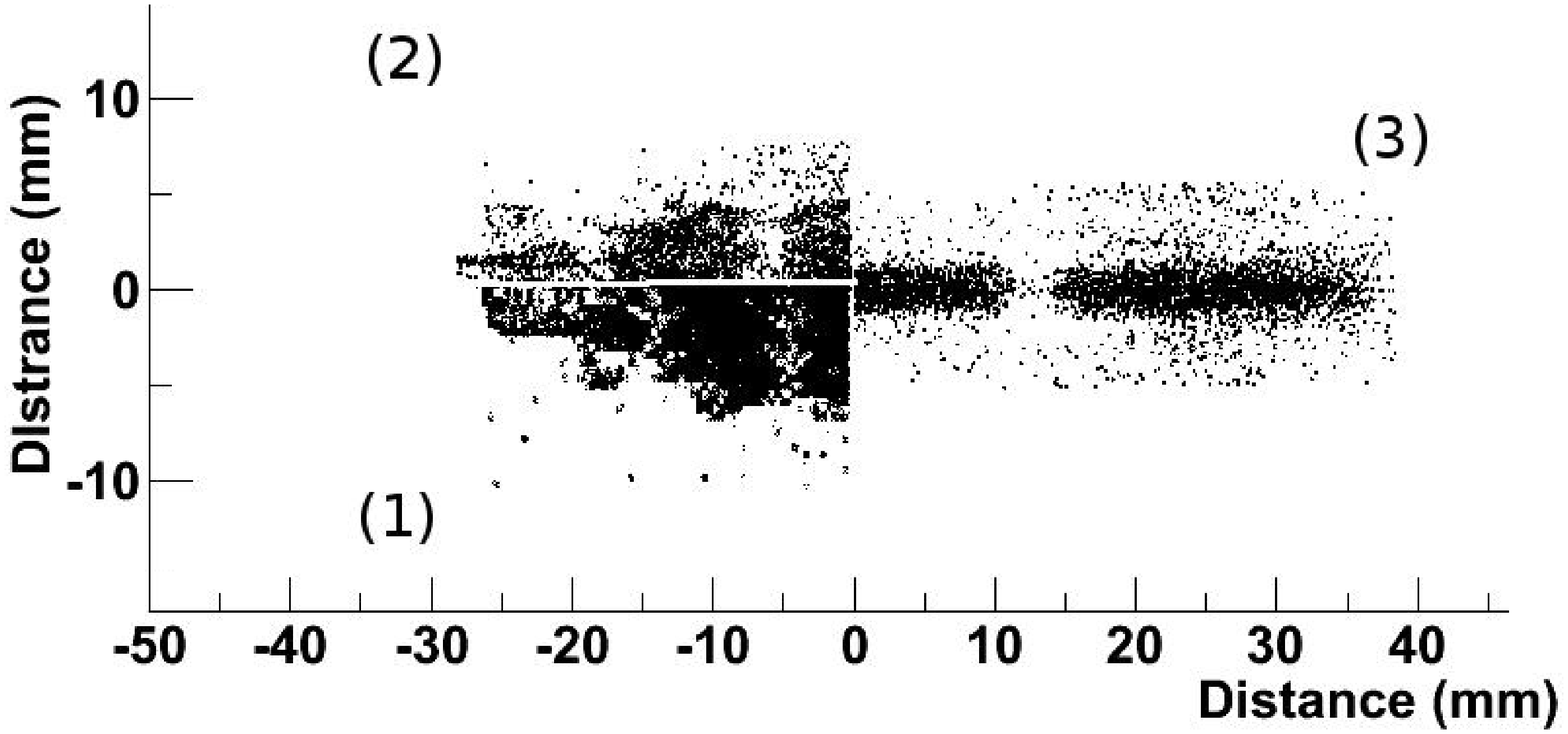}
\par\end{centering}

\caption{(a) Transport of the beam through 3 quadrupoles at 7.5 MeV using three
RCF segments placed at 296mm(1), 372mm(2) , and 620mm(3) from the
scatter foil. (b) Matching GEANT4 simulation run using 2x10$^{7}$
MeV protons and matching RCF positions.\label{fig:transport_triplet_7_5}}

\end{figure}

Using a triplet of quadrupoles in A-B-A configuration, where B is
an equivalent quadrupole to A rotated through 90\textdegree{} around
the beamline direction, gives a distinct focus point in one plane
(the plane of the focussing direction of both {}``A'' magnets).
Such a system can transport a proton beam with very little dispersion
in the focussing plane, with the caveat that a larger deegree of dispersion
will occur in the defocussing plane. In this experiment, a different
set-up (See Fig~\ref{fig:The-pepperpot_realsetup}) was used with
sections of RCF placed at intervals downstream from the magnet system.
It is clear in Fig~\ref{fig:transport_triplet_9} that the triplet
system has highly constrained a divergent beam with 14\textdegree{}
opening angle and 5.9 \textpm{} 0.2 MeV energy in the focussing plane.
This effect is even stronger when the beam energy is 3.8 \textpm{}
0.2 MeV with a slightly different opening angle of 16\textdegree{},
as can be seen in Fig~\ref{fig:transport_triplet_7_5}. 

\begin{figure}[H]
\begin{centering}
\includegraphics[width=0.75\paperwidth]{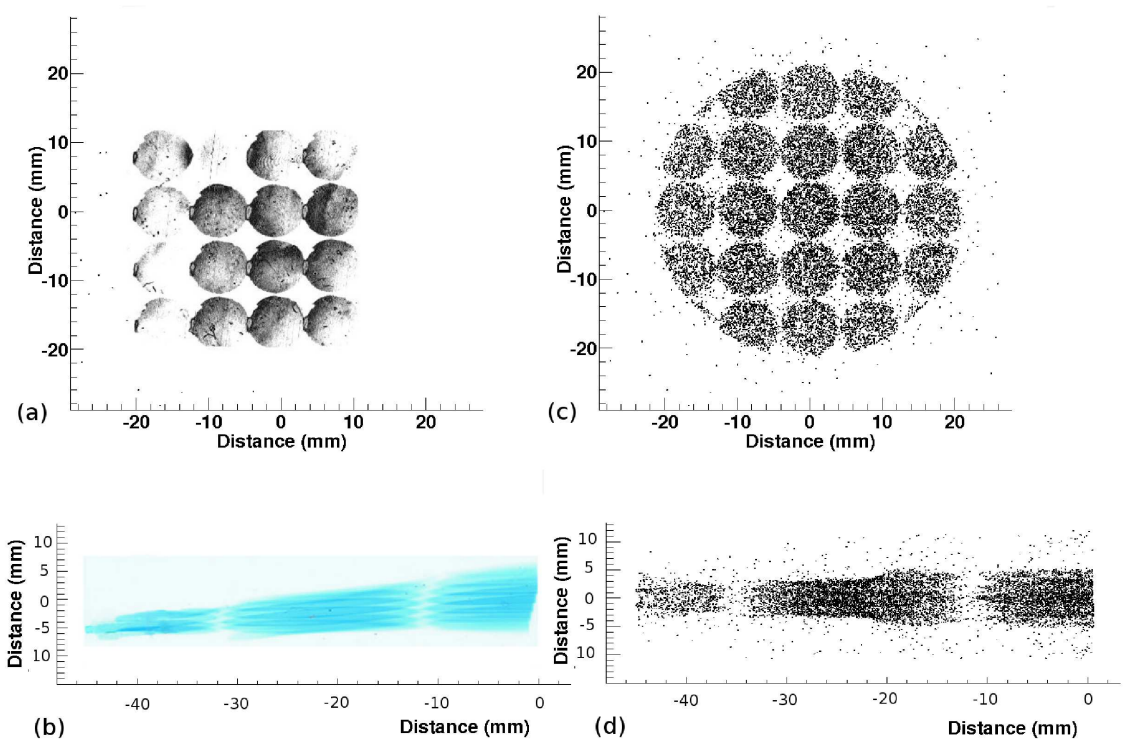}
\par\end{centering}

\caption{A comparison of RCF results with no magnets present at 442mm from
the scatter foil(a), and with the PMQ triplet present at 620mm from
the scatter foil(b). This is reproducible with GEANT4 (see (c) and
(d)).\label{fig:transport_analysis_1}}

\end{figure}

A comparison of the beam distribution at 442mm from the scatter foil
without the PMQ triplet, and at 620mm with the triplet in place, as
shown in Fig~\ref{fig:transport_analysis_1}, shows very clearly
that Halbach design PMQs can dramatically shape and intensify divergent
particle beams, and therefore should be extremely useful in many forthcoming
applications of high power laser acceleration technology.

\begin{figure}[H]
\begin{centering}
(a)\includegraphics[width=0.35\paperwidth]{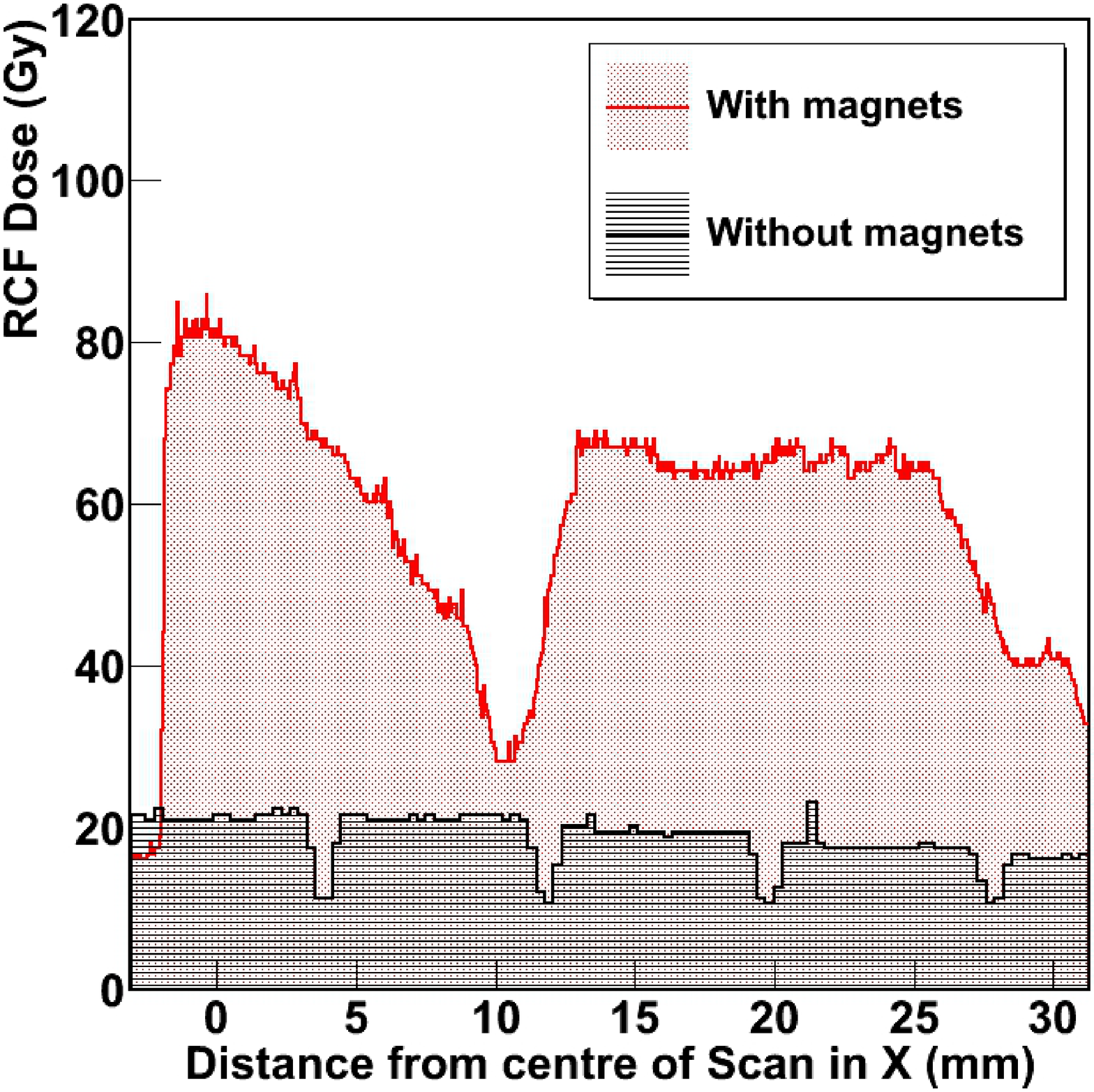}(b)\includegraphics[width=0.35\paperwidth]{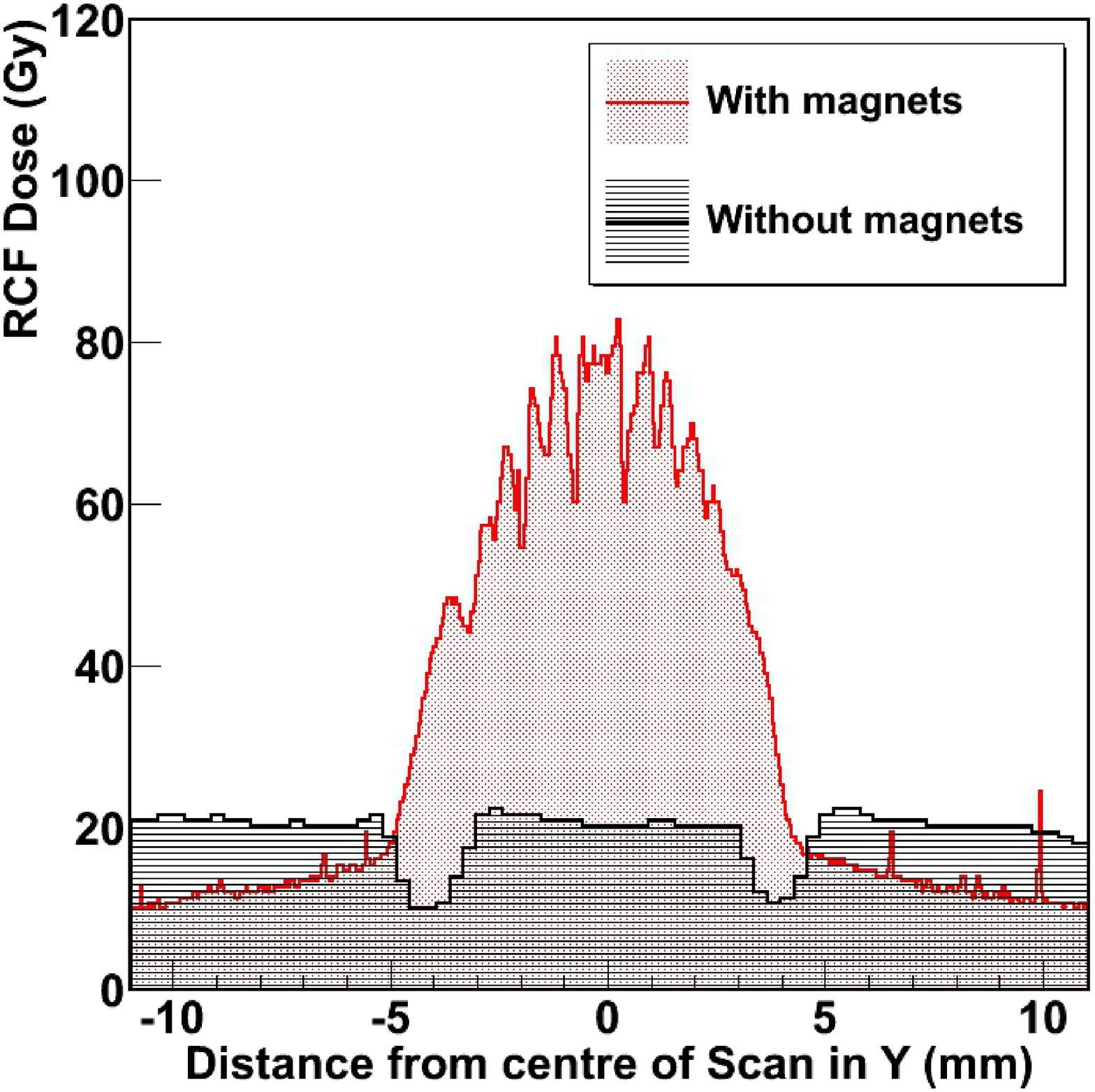}
\par\end{centering}

\caption{Analysis of the dose measured in the RCFs shown in Fig~\ref{fig:transport_analysis_1}(a)
and (b). Note that the orientation of Fig~\ref{fig:transport_analysis_1}(b)
was rotated 180\textdegree{} for the purpose of this analysis since
a re-scan was required to analyse the dose. A profile analysis is
done in the horizontal direction (a) and vertical direction (b). Red
shows data obtained using the PMQ triplet, and Black shows data obtained
with no magnets in the beamline.\label{fig:RCF_triplet_analysis}}

\end{figure}

An analysis of the PMQ triplet RCF data shown in Fig~\ref{fig:transport_analysis_1}
was carried out using the calibrated dose routine in ROOT, with the
results shown in Fig~\ref{fig:RCF_triplet_analysis}. The increase
in intensity seen on the RCF film using the PMQ triplet rises as high
as 400\% at peak. Allowing for the fact that the data with no PMQ
triplet present was taken closer to the scatter foil by 178 mm, it
is expected that a future experiment would demonstrate a slightly
higher increase in intensity if the experiment was repeated at the
same distance in both cases.

\begin{figure}[H]
\begin{centering}
(a)\includegraphics[width=0.32\paperwidth]{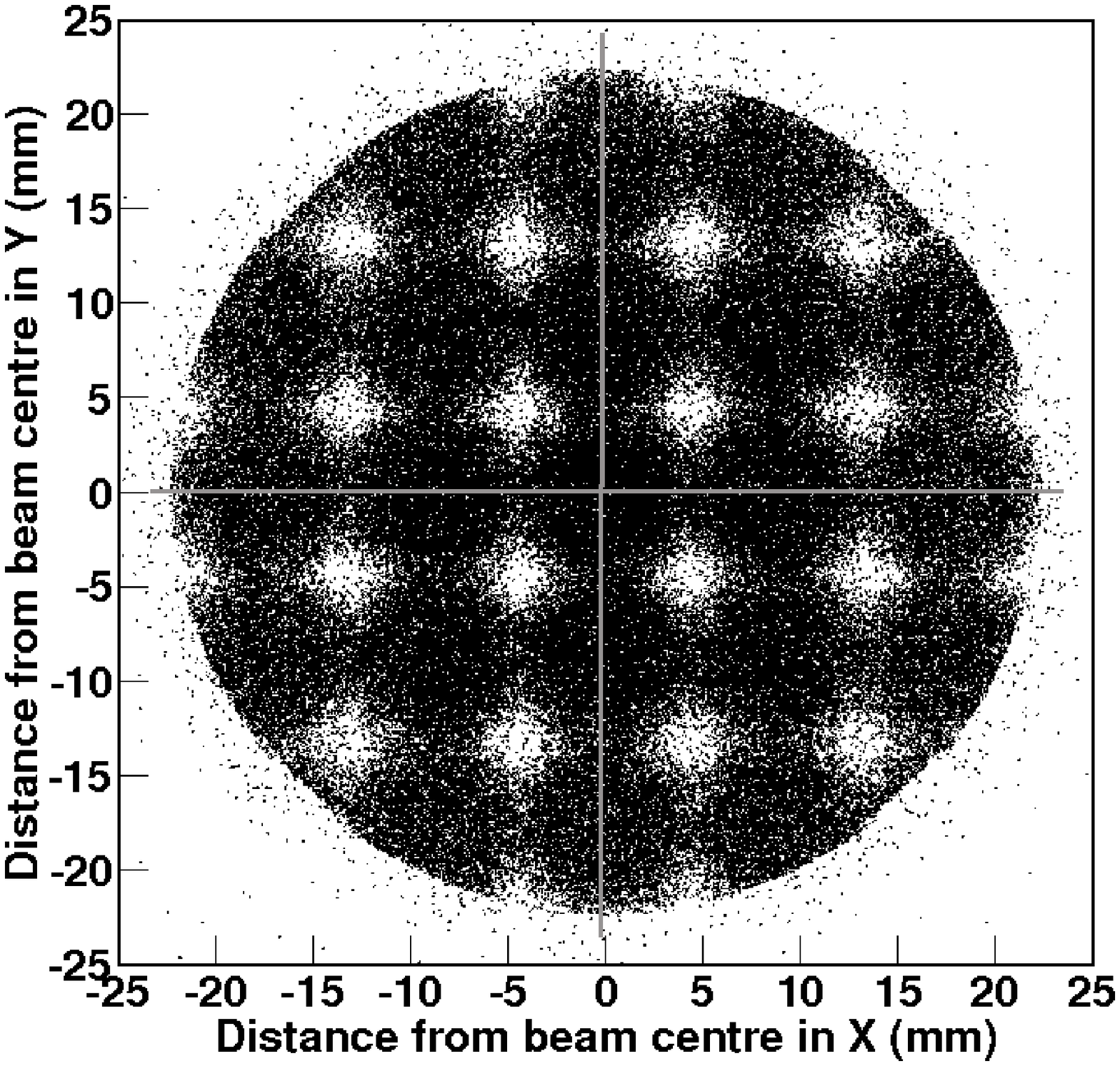}(c)\includegraphics[width=0.35\paperwidth]{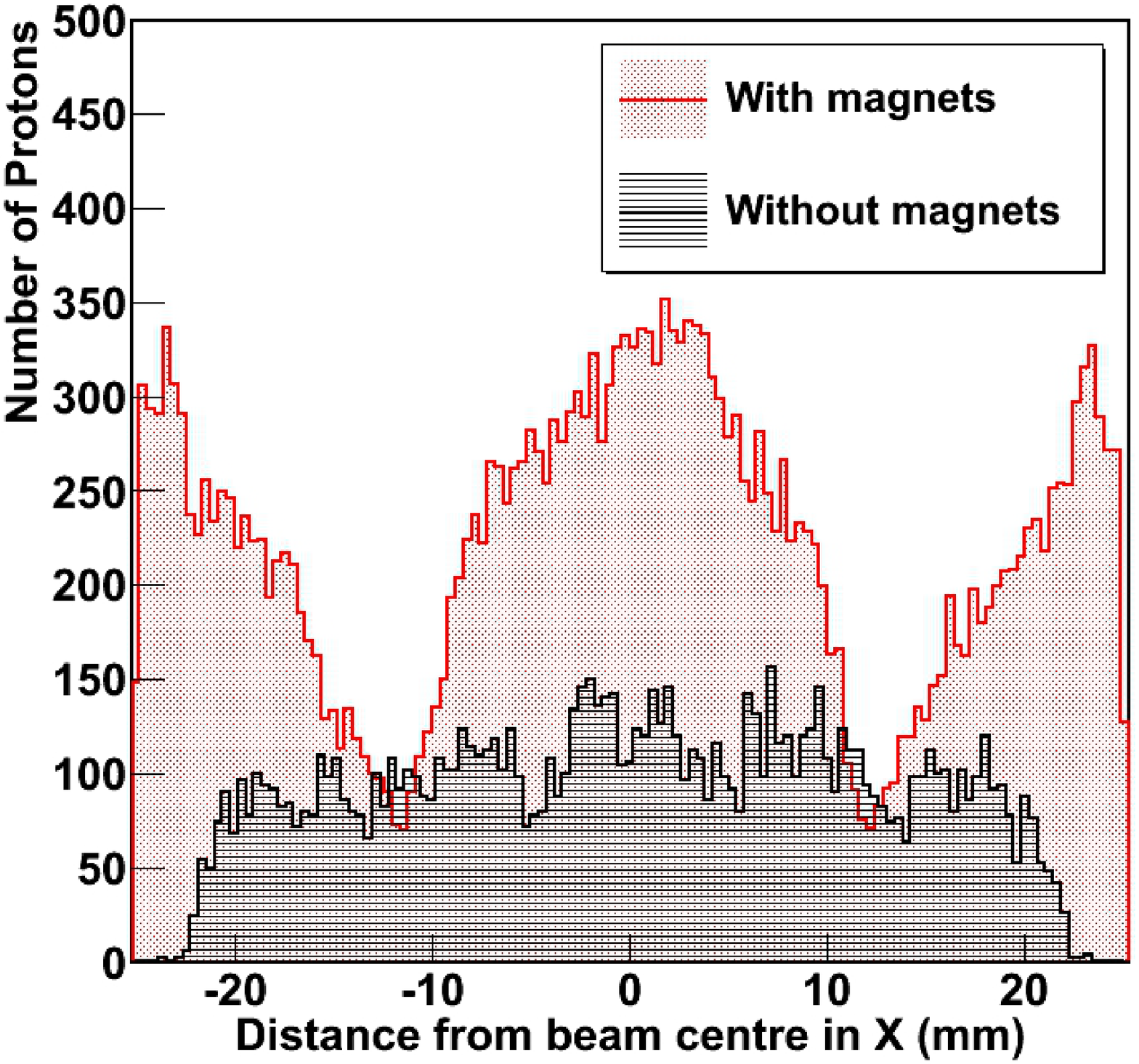}
\par\end{centering}

\begin{centering}
(b)\includegraphics[width=0.32\paperwidth]{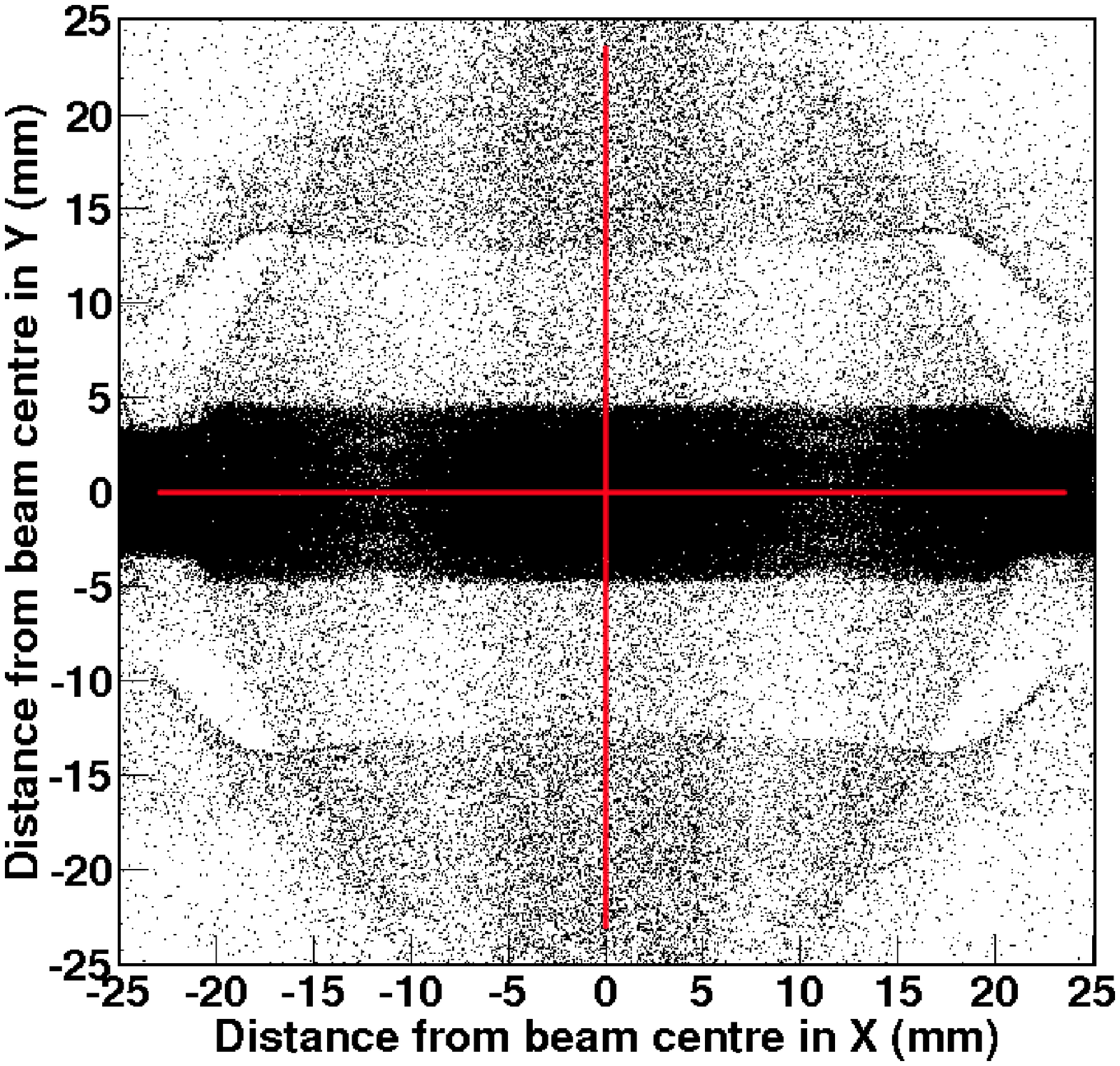}(d)\includegraphics[width=0.35\paperwidth]{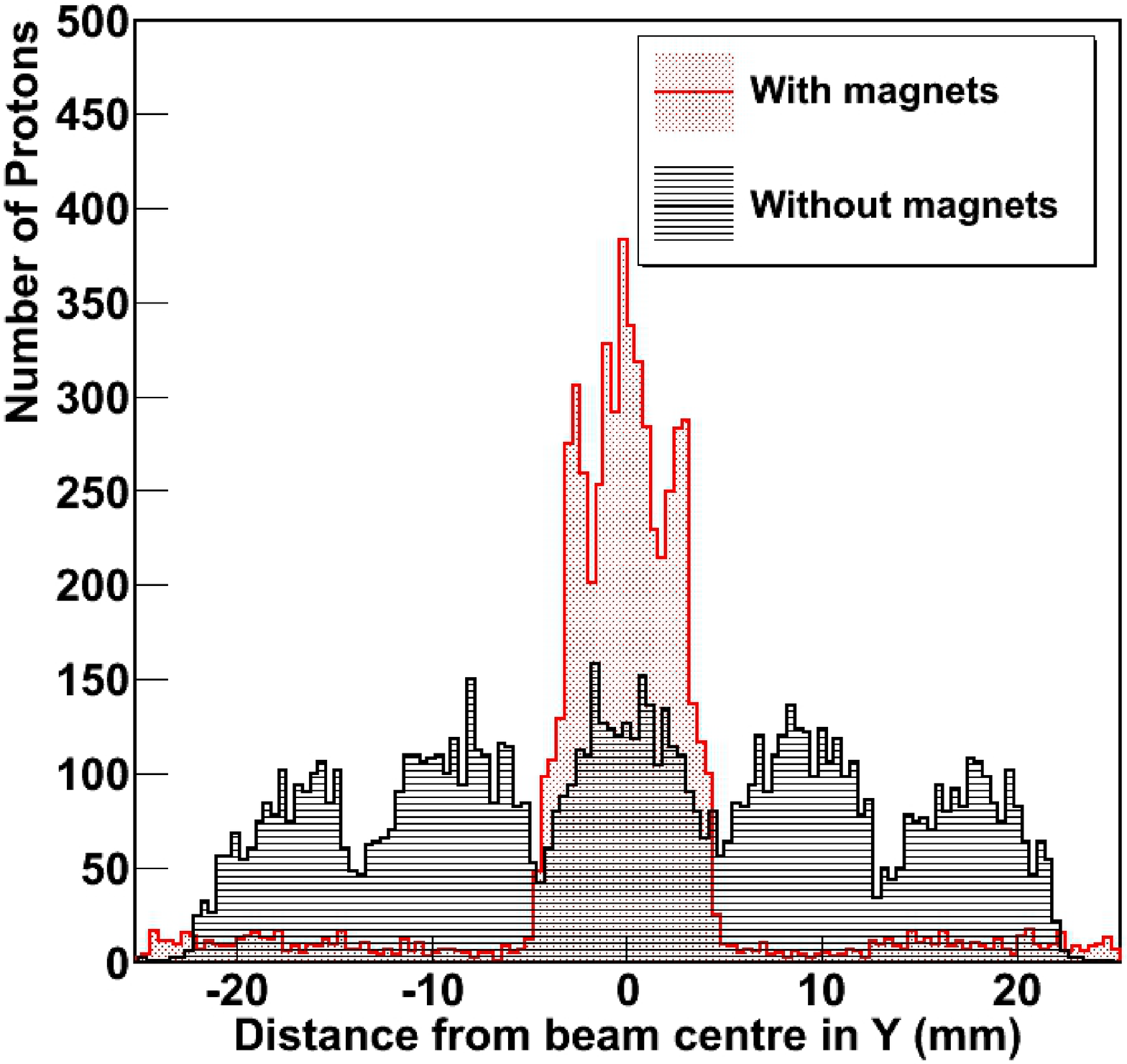}
\par\end{centering}

\caption{A projection analysis in GEANT4 can quantify the degree to which the
beamlets are focussed in a single plane, and the changes in angular
acceptance become clear. Comparing between the simulated data with
and without the PMQ triplet in place, the analysis is shown for the
horizontal plane in (c), and the same analysis performed in the vertical
plane is shown in (d). The vertical and horizontal lines on (a) and
(b) indicate the position of the analysis. In all of the above the
simulation results obtained with the PMQs in place are shaded red,
and the equivalent results with no PMQS are shaded black.\label{fig:transport_analysis_2}}

\end{figure}

Fig~\ref{fig:transport_analysis_2} shows an analysis done in GEANT4
of the proton distribution at 442mm behind the scatter foil, with
no PMQ triplet in place and at 620mm behind the scatter foil with
the PMQ triplet in place. The intensity of the proton beam, as expected,
increases dramatically in the focus plane, with up to a four-fold
increase in radiation per unit area in the vertical plane, which agrees
closely with the real data in Fig~\ref{fig:RCF_triplet_analysis}.

The relative intensities predicted by GEANT4, with and without the
PMQs present, are in agreement with those seen in the experimental
data. One clearly visible effect not accounted for in the simulation
is the overlapping of pepperpot beamlets on the RCF film which can
be seen clearly in Fig~\ref{fig:Pepperpot-results-real}(b), which
could be due to the far greater number of protons produced during
each experimental run than it is possible to simulate in a reasonable
time frame with small-scale computational resources. It is expected
that this discrepancy could be reduced to some extent by greatly increasing
the computational time used for the GEANT4 simulations.

\section{Conclusion}

The methods described in this paper adequately demonstrate the focussing
characteristics of a Permanent Magnetic Quadrupole system for both
a pencil beam and a diverging beam, and clearly show that such a system
will be very useful for a compact proton/ion transport and focussing
apparatus. With the advent of new facilities producing laser-accelerated
beams of protons and ions, such techniques will be of immense importance
for control and optimisation of these beams for physics and medical
applications. It is envisaged that advancements on this method, especially
increasing the strength and number of the PMQs, will allow new avenues
of research in nuclear and plasma physics of astrophysical relevance.
When coupled with dipoles to disperse proton energies this technology
can also be used to provide mono-energetic beams from the broad energy
range of beams typically seen with laser produced protons. Using the
studies and simulation techniques presented here, we hope to use these
techniques on a laser system to demonstrate the inherent benefits
that focussing, control and transport of a laser-produced proton beam
have in comparison with current accelerator technology. 

%\bibliographystyle{unsrtnat}
%\bibliography{refs_pmq_v2}

\end{document}